\documentclass[a4paper,fleqn,usenatbib]{mnras}

\usepackage{newtxtext,newtxmath}
\usepackage[T1]{fontenc}
\usepackage{ae,aecompl}
\usepackage{comment}
\usepackage[nolist,nohyperlinks,printonlyused]{acronym}
\usepackage{subfigure}

\usepackage{float}

\usepackage{graphicx}	% Including figure files
\usepackage{amsmath}	% Advanced maths commands

\newcommand{\HI}{\ion{H}{I}}

\newcommand{\mathHI}{{\mbox{\scriptsize \HI}}}

\newcommand\given[1][]{\:#1\vert\:}

\newcommand{\lyaf}{\text{Ly$\alpha$ forest}}
\newcommand{\lya}{\text{Ly$\alpha$}}

\newcommand{\NHI}{$N_{\mathHI{}}$}

\newcommand{\NHIt}{N_{\mathHI{}} }

\newcommand{\vpfit}{\texttt{VPFIT}}
\newcommand{\bndist}{$b$-$N_{\mathHI{}}$~distribution}
\newcommand{\bn}{$\left\{b, N_{\mathHI{}}\right\}$}
\newcommand{\dndz}{d$N$/d$z$}

\newcommand{\bnpdfG}{$P ( b \mathbin{,} N_{\mathHI{}} \, \given \, T_0, \gamma , \Gamma_{\mathHI{}})$}

\title[Measuring the IGM thermal and ionization state]
{Measuring the thermal and ionization state of the low-$z$ IGM using likelihood free inference}

\author[Hu et al.]{
Teng Hu,$^{1}$\thanks{E-mail: tenghu@ucsb.edu (UCSB)}
Vikram Khaire$^{1,2}$,
Joseph F. Hennawi$^{1,3}$,
Michael Walther$^{1,4}$,
Hector Hiss$^{5}$,\newauthor
Justin Alsing$^{6,7}$,
Jose O\~norbe$^{8}$,
Zarija Lukic$^{9}$ and
Frederick Davies$^{5}$
\\
$^{1}$Physics Department, Broida Hall, University of California Santa Barbara, Santa Barbara, CA
93106-9530, USA\\
$^{2}$ Indian Institute of Space Science \& Technology, Thiruvananthapuram, Kerala - 695547, INDIA\\
$^{3}$Leiden Observatory, Leiden University, PO Box 9513, NL-2300 RA Leiden, the Netherlands\\
$^{4}$University Observatory, Faculty of Physics, Ludwig-Maximilians-Universität München, Scheinerstr. 1, 81679 Munich, Germany\\
$^{5}$Max-Planck-Institut für Astronomie, Königstuhl 17,69117 Heidelberg, Germany\\
$^{6}$Oskar Klein Centre for Cosmoparticle Physics, Department of Physics, Stockholm University, Stockholm SE-106 91, Sweden\\
$^{7}$Imperial Centre for Inference and Cosmology, Department of Physics, Imperial College London, Blackett Laboratory,Prince Consort Road, London SW7 2AZ, UK\\
$^{8}$Facultad de F\'isica, Universidad de Sevilla, Avda. Reina Mercedes s/n, Campus de Reina Mercedes, E-41012 Sevilla, Spain\\
$^{9}$Lawrence Berkeley National Laboratory, Berkeley, CA 94720, USA\\
}

\date{Accepted XXX. Received YYY; in original form ZZZ}

\pubyear{2021}

\begin{document}
\label{firstpage}
\pagerange{\pageref{firstpage}--\pageref{lastpage}}
\maketitle

\begin{abstract}
We present a new approach to measure the power-law temperature density relationship $T=T_0 (\rho \slash \bar{\rho})^{\gamma -1}$ and the UV background photoionization rate $\Gamma_{\mathHI{}}$ of the \ac{IGM} based on
the Voigt profile decomposition of the \lya{} forest into a set of discrete absorption lines with Doppler parameter $b$ and the neutral hydrogen column density $N_{\rm HI}$. Previous work demonstrated that the shape of the \bndist{} is sensitive
to the IGM thermal parameters $T_0$ and $\gamma$, whereas our new inference algorithm also takes into account the normalization of the distribution, i.e. the line-density \dndz{}, and we demonstrate that precise constraints can also be obtained on  $\Gamma_{\mathHI{}}$. We use density-estimation likelihood-free inference (DELFI) 
to emulate the dependence of the \bndist{} on IGM parameters trained on an ensemble of 624 Nyx hydrodynamical simulations at $z = 0.1$, which we combine with a Gaussian process emulator of the normalization. To demonstrate the efficacy of this approach, we generate hundreds of realizations of realistic mock HST/COS datasets, each comprising 34 quasar sightlines, and forward model the noise and resolution to match the real data. We use this large ensemble of mocks to extensively test our inference and empirically demonstrate that our posterior distributions are robust. 
Our analysis shows that by applying our new approach to 
existing Ly$\alpha$ forest spectra at $z\simeq 0.1$, one can measure the thermal and ionization state of the \ac{IGM} with
very high precision ($\sigma_{\log T_0} \sim 0.08$ dex, $\sigma_\gamma \sim 0.06$, and $\sigma_{\log \Gamma_{\mathHI{}}} \sim 0.07$ dex).
\end{abstract}


%%%%%%%%%%%%%%%%%%%%%%%%%%%%%%%%%%%%%%%%%%%%%%%%%%

%%%%%%%%%%%%%%%%% BODY OF PAPER %%%%%%%%%%%%%%%%%%
\section{Introduction}
  \label{sec:intro}

The intergalactic medium (\ac{IGM}) is the largest reservoir of baryons in the Universe, 
which plays an essential role in its evolution and structure formation. 
Current theoretical models,
supported by many observations,
predict two major phase transition events that dominate the thermal evolution of the \ac{IGM}. 
The first one is the reionization of hydrogen by the first galaxies at redshift $6 < z < 20$ \citep{Madau1998,Faucher-Giguere2008,Robertson2015,mcgreer1,Fan2006}. 
The second phase transition is the double reionization of Helium
(\ion{He}{II}$\rightarrow$\ion{He}{III}) driven by \ac{QSO}s \citep[see e.g.][]{MadauMeiksin1994, MiraldaEscude2000, mcquinn09, Dixon2009, Syphers2014}, 
which is expected to happen at $z \sim 3$, where the quasar luminosity density peaks
\citep[see e.g.][]{Worseck2011,Khaire2017, Worseck2018,Kulkarni2019}. 
These two events change the ionization state of the \ac{IGM} dramatically and 
heat it to temperatures as high as 15,000K.

After hydrogen reionization ($z<6$),
the thermal state of the \ac{IGM} is determined by the 
balance between photoionization heating from the extragalactic UV background 
and various cooling processes such as  
cooling due to Hubble expansion, recombinations, and the excitation and inverse Compton scattering of electrons from the cosmic microwave background (CMB).  
As a result of these processes, the \ac{IGM} is expected to follow a 
tight temperature-density relation: 
\begin{equation}
T (\Delta) = T_0 \Delta^{ \gamma -1},
\label{eqn:rho_T}
\end{equation}
where $\Delta = \rho/ \bar{\rho}$ is the overdensity, $T_0$ is the temperature at mean density,
and $\gamma$ is the adiabatic index \citep{hui1, McQuinn2016},
and these two parameters characterize the thermal state of the \ac{IGM}.
By measuring $T_0$ and $\gamma$ at different epochs, 
we are thus able to constrain the \ac{IGM} thermal history \citep{Miralda1994,Hui2003},
improving our knowledge of the evolution of the \ac{IGM} and our understanding of the relevant heating and cooling processes responsible.

The thermal state of the IGM is encoded in the \lyaf, 
a swath of \lya{} absorption 
lines originating from a trace amount of neutral hydrogen gas in the IGM 
\citep{GunnPeterson1965, lynds}. The \lyaf{} is thus used as the premier probe of the 
\ac{IGM} thermal history. 
Various statistical properties of the \lyaf{} are used to measure the \ac{IGM} thermal 
state, including the power spectrum \citep[][]{Theuns2000,Zaldarriaga2001,McDonald2001,Walther2017,Walther2018,Khaire2019,Gaikwad2021},
the flux probability density function (PDF) \citep[][]{Bolton2008,Viel2009,Lee2015}, 
the transmission curvature \citep[][]{Becker2011,Boera2014}, 
the wavelet decomposition of the forest \citep[][]{Theuns2000b,Theuns2002,Lidz2010,Garzilli2012,Wolfson2021},
and the quasar pair phase angle distribution \citep[][]{Rorai2013,Rorai2017}.
These measurements are typically performed using \lyaf~spectra 
from ground-based telescopes at $z > 1.6$,  
where the \lya{} transition lies above the atmospheric cutoff ($\lambda \sim 3300 \text{\AA{}}$), 
explaining why there are currently very few measurements of the \ac{IGM} thermal state at
redshift below such limit (i.e. $z<1.6$),
which is, however, an essential epoch for galaxy formation.
By far the only available direct measurements at redshift $z<1.6$ is reported by \citet[][]{ricotti2000} at $z\sim 0$, which was done two decades ago using only 43 Ly-$\alpha$ absorption lines from HST Goddard High Resolution Spectrograph data,
suggesting a need for new and precise measurements at redshift $z \sim 0$.

Long after the helium reionization ($z < 3$), the thermal 
state of the IGM is expected to be dominated by adiabatic cooling from Hubble expansion, 
where theoretical models and simulations predict such cooling 
leads to an \ac{IGM} thermal state with $ T_0\sim 5000$K and $\gamma \sim 1.6$ 
at the current epoch $z=0$ \citep{McQuinn2016}. 
However, to date, this predicted cooling to 
low temperatures has not been verified observationally. 
Moreover, recent studies based on the low-$z$ \lyaf{} dataset \citep{Danforth2016} show that
these lines appear broader (i.e. have larger $b$ parameter) than numerical model predictions \citep{Gaikwad2017, Viel2017, Nasir2017}.
While it has been speculated that such a discrepancy might be resolved by an additional source of turbulence \citep{Bolton2021},
an alternative explanation would be that there are additional sources of heating, and the \ac{IGM} is actually hotter than expected, with $T_0$ conceivably approaching $10000$K.

If true, such unexpected heating would change our understanding of \ac{IGM} physics drastically,
highlighting a severe need to investigate processes that are possibly responsible for it,
such as dark matter annihilation \citep[][]{Araya2014}, 
gamma ray sources \citep[][]{Puchwein2012},
or feedback from galaxy formation, whose effects are not fully understood in low-$z$ \citep[see][]{Springel2005,Croton2006,Sijacki2007,Hopkins2008}.
To this end, precise measurements of the thermal state at low-$z$ are 
needed to determine whether the \ac{IGM} cools down as predicted.

In this work, we follow the method for measuring 
the \ac{IGM} thermal state based on Voigt profile decomposition
of the \lyaf{} \citep[][]{schaye1999,ricotti2000,McDonald2001}. 
In this approach, a transmission spectrum 
is treated as a superposition of multiple discrete Voigt profiles, 
with each line described by three parameters: redshift $z_{\text{abs}}$, Doppler broadening $b$, 
and neutral hydrogen column density $N_{\rm HI}$. 
By studying the statistical properties of these parameters, 
i.e. the \bndist{},
one can recover the thermal information encoded in the absorption profiles. 
The majority of past applications of this method constrained the IGM thermal state by fitting the low-$b$-$N_{\rm HI}$ cutoff of 
the \bndist{} \citep[][]{schaye1999,schaye2000,ricotti2000,McDonald2001,rudie2012,bolton2014,Boera2014,Garzilli2015,Garzilli2018, Rorai2018,Hiss2018}. 
The motivation for this approach is that the \lya{} lines are broadened 
by both thermal motion and non-thermal broadening
resulting from  combinations of Hubble flow, peculiar velocities and turbulence.
By isolating the narrow lines in the \lyaf{}
that constitutes the lower-cutoff in \bndist{}s, 
of which the line-of-sight component 
of non-thermal broadening
is expected to be zero, 
the broadening should be purely thermal, 
thus allowing one to constrain the IGM thermal state.  
However, this method has three crucial drawbacks.  
First, the \ac{IGM} thermal state actually impacts all the lines besides just the narrowest lines.
Therefore, by restricting attention to data in the distribution outskirts, 
this approach throws away information and reduces the sensitivity to the IGM thermal state significantly\citep{Rorai2018,Hiss2019}.
Second, in practice, determining the location of the cutoff is vulnerable to systematic effects,
such as contamination from the narrow metal lines \citep[][]{Rorai2018,Hiss2018}. 
Lastly, the results from this approach critically depend on the choice of low-$b$ cutoff fitting techniques,
where different techniques might result in inconsistent $T_0$ and $\gamma$ measurements \citep[][]{Rorai2018,Hiss2018}.

To overcome these limitations, \citet[][]{Hiss2019} developed a new approach to measure the \ac{IGM} thermal state from the full \bndist{} based on density estimation and Bayesian analysis. 
We further advance the \bndist{} emulation by employing a novel density estimation technique based on machine learning, namely Density-Estimation Likelihood-Free Inference (DELFI) 
\citep[see][]{papamakarios2016, Alsing2018, papamakarios2018, Lueckmann2018, Alsing2019}. 
In addition, we augment the likelihood function to take into account the absorber number density \dndz{},
making our improved method far more sensitive to the photoionization rate of hydrogen $\Gamma_{\mathHI{}}$ sourced by the UV background.

In this work, we introduce our new method, demonstrate its robustness, 
and perform an analysis using realistic mock datasets to illustrate the sensitivity to IGM parameters. 
Our inference is based on a suite of cosmological hydrodynamic simulations with different thermal parameters at redshift $z \sim 0.1$.
While this method can be applied to the Ly$\alpha$ forest at any redshift where the opacity 
is low enough to make it amenable to Voigt profile decomposition \citep[e.g. $z\lesssim 3.4$, see][]{Hiss2018}, 
we choose to focus on $z\sim 0.1$ because we want to quantify the sensitivity of archival \emph{Hubble Space Telescope} spectra, 
so as to perform the first measurements of the \ac{IGM} thermal state at $z < 1.6$ in future work.
Such a measurement would directly test the prediction that the IGM cools down
at low-$z$, which has been challenged by recent observations.
To this end, we run a set of cosmological hydrodynamic simulations with 
different thermal parameters at redshift $z \sim 0.1$, 
from which we create mock datasets with the same properties as 
the \citet{Danforth2016} low redshift Ly$\alpha$ forest dataset observed with the \emph{Cosmic Origins Spectrograph} \citep[COS,][]{Green2012} on the \ac{HST}. 
We demonstrate that our method applied to such a dataset can reliably and accurately determine the thermal state of the IGM.

This paper is structured as follows. 
In \S\ref{3sec:simulations} we introduce our hydrodynamic simulations,
parameter grid, and data processing procedures, 
which include generating \lyaf{} from simulation,
forward-modeling and our method to fit Voigt profiles (\vpfit{}). 
In \S\ref{sec:inference} we 
present our inference algorithm, including likelihood, emulators, inference results,
and a set of inference tests. 
Finally, we discuss these results and summarize the highlights of this study in 
\S\ref{3sec:discussion}. 
Throughout this paper, we write  $\log$ in place of $\log_{10}$.
Cosmology parameters used in this study 
($\Omega_m = 0.319181, \Omega_b h^2 = 0.022312, h= 0.670386, n_s = 0.96, \sigma_8 = 0.8288$) are taken from \citet{Planck2014} .

 \begin{figure*}
 \centering
    \includegraphics[width=0.99\linewidth]{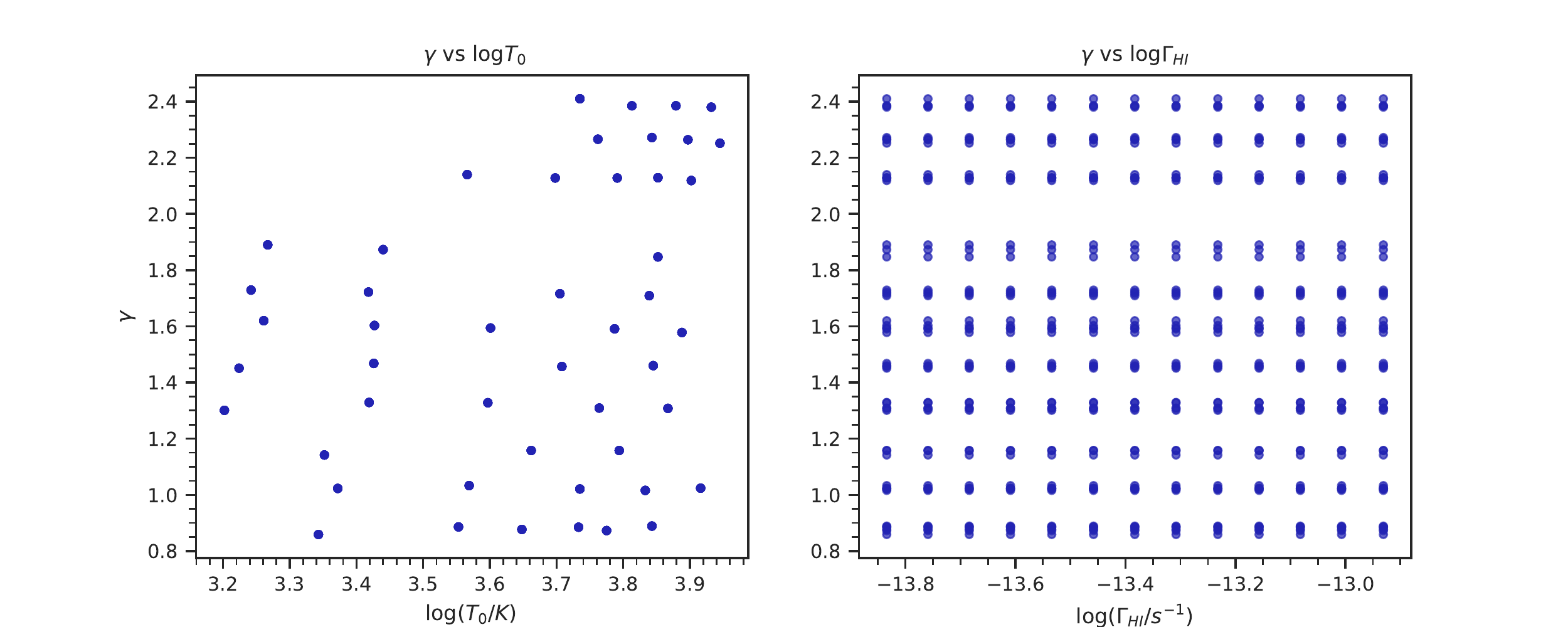}
  \caption{Thermal grid (blue circles) from snapshots of hydrodynamic simulations of the \ac{THERMAL} suite at $z=0.1$. 
  The left-hand panel is the $\gamma$ - $T_0$ grid, 
  whose shape is determined by the parameters of thermal grid at and
  the evolution of the thermal state of the IGM. 
  The right-hand panel is $\gamma$ - $\Gamma_{\mathHI{}}$ grid,
  showing the thirteen $\Gamma_{\mathHI{}}$ values for each point on the 2D $\gamma$ - $T_0$ grid. }
  \label{3fig:lowz_grid}
\end{figure*} 
 
 \section{Simulations}
  \label{3sec:simulations}

 \begin{figure*}
 \centering
\includegraphics[width=0.49\textwidth]{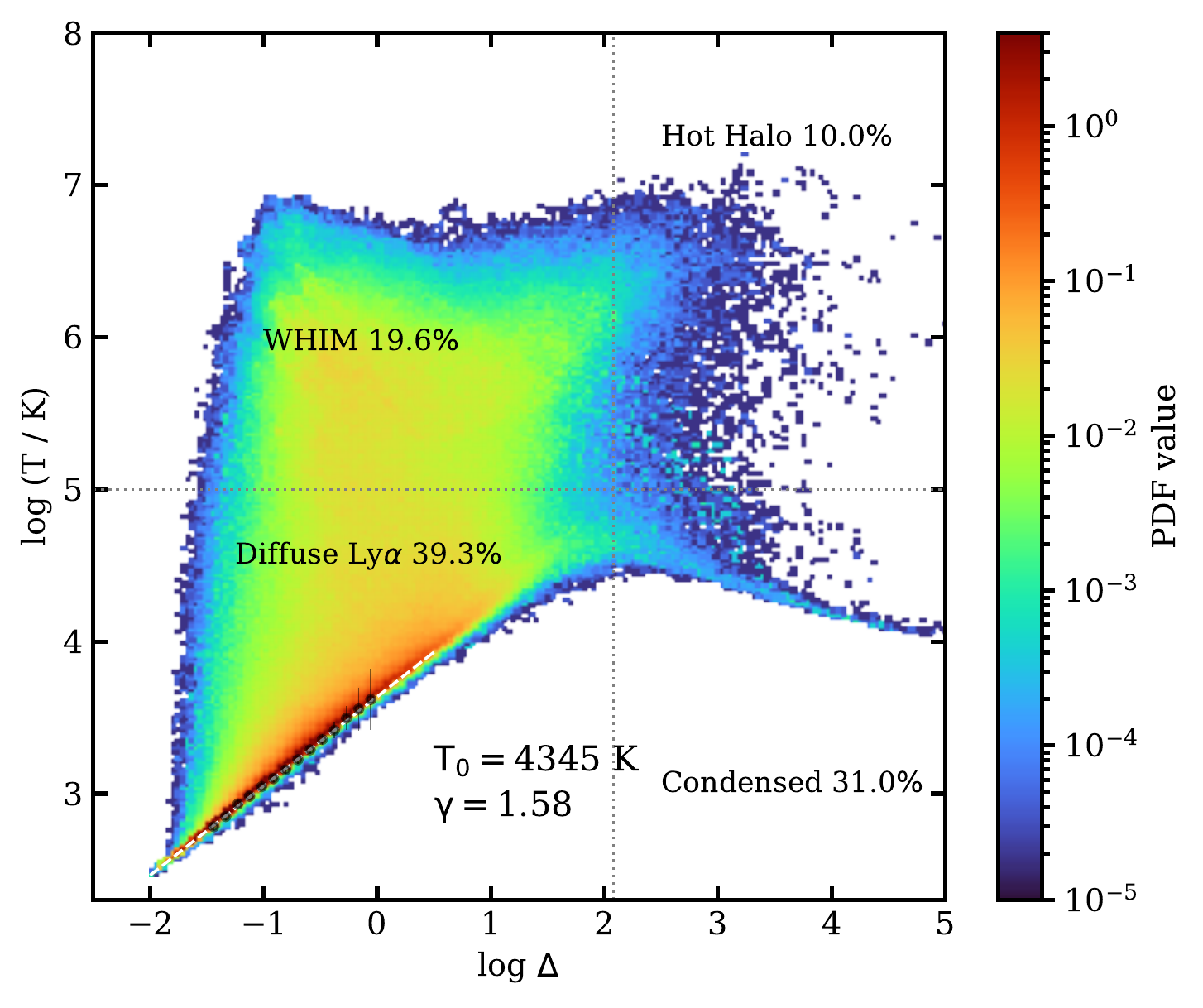}
\includegraphics[width=0.49\textwidth]{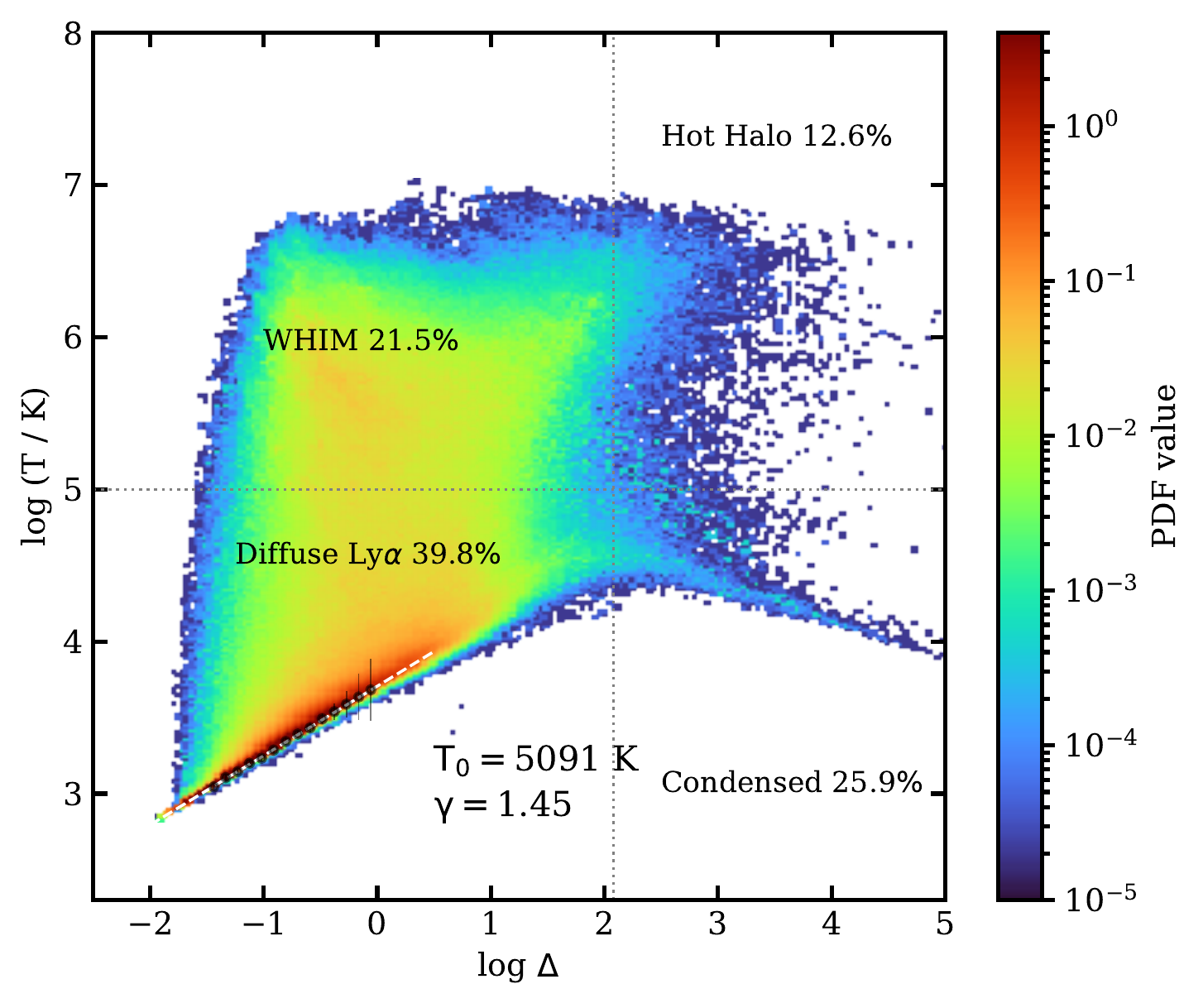}
  \caption{Temperature-density ($T$-$\Delta$) distribution for the IGM gas in two different models from Nyx simulation. 
  White dashed lines is the power-law fit to the $T$-$\Delta$ relation,
  and legends show the best fit values of $T_0$ and $\gamma$. 
  Dotted T = $10^5$ K lines divide the phase diagram into hot and cold region, while only cold gas is used for the fitting. 
  Density peaks ($\log T_\text{peak,i}$, $\log\Delta_i$) for each bin are plotted as black dot, and 1-$\sigma_{T,i}$ error bars are shown as black bars. The left-hand panel shows a model with $T_0=4345$ K and $\gamma=1.58$,
  and the right-hand panel shows a model with $T_0=5091$ K and $\gamma=1.45$. The density weighted gas phase fractions are shown in the annotation.}
  \label{3fig:Nyx_Rho_T}
\end{figure*} 

A set of Nyx cosmological hydrodynamic simulations \citep[see][]{Lukic2015, Almgren2013} is used to model the low-redshift \ac{IGM}.
Nyx is a massively-parallel, cosmological simulation code primarily developed to simulate the \ac{IGM}. \footnote{
Nyx simulation is able to run with
Adaptive Mesh Refinement (AMR). However, the AMR feature is not used in this work, 
since this work focus on the \lyaf{}, 
which distribute nearly the entire simulation domain rather than isolated concentrations of matter where AMR is more effective.
}
In Nyx simulations, the evolution of dark matter is traced by treating dark matter 
as self gravitating Lagrangian particles, while baryons are modeled as as ideal gas on a
uniform Cartesian grid following an Eulerian approach. The Eulerian gas dynamics equations 
are solved following a second-order piece-wise parabolic method (PPM),
which captures shock waves accurately.

Nyx includes the main physical processes relevant for modeling the \lyaf. 
First of all, gas in the Nyx is assumed to have a primordial composition with a hydrogen mass fraction of 0.76,
and helium mass fraction of 0.24 and zero metallicity. 
The recombination, collisional ionization, dielectric recombination,
and cooling are implemented based on prescriptions given in \citet{Lukic2015}. 
Nyx keeps track of the net loss of thermal energy resulting from atomic collisional processes and takes into account the inverse Compton cooling off the microwave background.
Ionizing radiation is modeled by a spatially homogeneous but time-varying ultraviolet background radiation field \citep[from][]{H&M2012} that changes with redshift, 
while assuming all cells in the simulation are optically thin. 
We later make the UV background a free parameter for generating \lya{} forest in post-processing (See \S\ref{sec:skewers}).
Since Nyx simulations are developed mainly to study the IGM, no feedback or galaxy formation processes are included, 
significantly reducing the computational requirement allowing us to run a large ensemble of simulations 
varying the thermal parameters
(see \ref{sec:parameter}).

Each Nyx simulation used in this study is initialized at $z=159$ 
and evolves down to $z=0.03$ in a $L_{\text{box}} = 20~{\rm cMpc}\slash h$ simulation domain,
using $N_{\text{cell}} = 1024^3$ Eulerian cells and $1024^3$ dark matter particles. 
The box size is chosen as the best compromise between computational cost 
and the need to be converged at least to $< 10\%$ on small scales (large $k$).
More discussion about resolutions and box sizes can be found in \citet{Lukic2015}. 
We also performed box size convergence tests at low redshift as explained in appendix~\ref{sec:convergence}.

 \subsection{Thermal parameters and simulation grid}
   \label{sec:Thermal para}

To model the \ac{IGM} with different thermal states, we use part of the publicly available Thermal History and Evolution in Reionization Models of Absorption Lines (\ac{THERMAL})\footnote{For details of THERMAL suite, see http://thermal.joseonorbe.com} suite of Nyx simulations \citep[see][]{Hiss2018,WaltherM2019}. 
We make use of in total 48 models with different thermal histories,
and for each model,
we generate a simulation snapshot at $z= 0.1$,
from which we measure the thermal state [$\log T_0$,$\gamma$].  
The thermal grid is illustrated in the left panel of Fig.\ref{3fig:lowz_grid}, 
which shows that $\log (T_0/\text{K})$  spans from $3.2$ to $3.95$,
and $\gamma$ ranges from $0.86$ to $2.41$.  
Here different thermal histories are achieved 
by artificially changing the photoheating rates ($\epsilon$)
following the method presented in \cite{Becker2011}. 
In this method, $\epsilon$ is treated as a function of overdensity, i.e. 
\begin{equation}
\epsilon =\epsilon_{\rm HM12} A \Delta^B, 
\label{eqn:heating}
\end{equation}
where  $\epsilon_{\rm HM12}$ represents the photoheating rate per ion tabulated in 
\cite{H&M2012}, and $A$ and $B$ are parameters that are varied to obtain 
different thermal histories. 
It is noteworthy that the thermal state tends to converge towards low redshifts due 
to the cooling and other physical processes in the evolution, 
and it is therefore difficult to generate models with a 
uniform grid of $T_0$ and $\gamma$ (for more details, see \citealt{WaltherM2019}). 
Moreover, it is especially challenging to generate models with
low $T_0 (< 10^{3.5}~{\rm K})$ and high $\gamma ( > 1.9)$ at low-$z$, 
because when one reduces the photoheating rates to obtain lower $T_0$, 
the cooling rate from Hubble expansion dominates,
and $\gamma$ asymptotically approaches values near 1.6 \citep[see][]{McQuinn2016}. 
As a result, the $T_0$-$\gamma$ grid has an irregular shape, 
and there are no models in the high $\gamma$ low $T_0$ regions. 
In addition, such an irregular $T_0$-$\gamma$ grid is also 
a result of the original grid of the \ac{THERMAL} suite,
which is driven by the high-$z$ thermal state analysis in \citet[][]{WaltherM2019} 
that obtains relatively high temperatures.

To measure the thermal state for each of the 48 models, 
we fit temperature-density ($T$-$\Delta$) relation (see Eq.~\ref{eqn:rho_T}) to the
temperatures and densities in the simulation domain.
While fitting the $T$-$\Delta$ relationship, 
we noticed broader distributions of the \ac{IGM} temperatures in low redshift ($z < 0.5$) compared to high redshift ($z>3$).
Examples of low-$z$ \ac{IGM} temperature-density distributions are
illustrated in Fig.~\ref{3fig:Nyx_Rho_T}, where we show 2D histograms of 
$T$-$\Delta$ of gas in each cell for two of our simulations on the {\sc thermal} grid at $z=0.1$. 
The gas cells are divided into four phases depending on the temperature and density, 
namely the \ac{WHIM}, Diffuse Ly$\alpha$, Hot Halo gas, and Condensed \footnote{Here we follow the definition used in \citet{Dave2010}, 
where the cutoffs are set to be $T=10^5$K and $\Delta$ = 120, more discussion about the different cutoff used in literature can be found in \citet{Gaikwad2017}.}.
The density-weighted gas phase fractions are shown in the legends of the figure,  
where the diffuse Ly$\alpha$ 
phase representing the densities and temperatures probed by the \lyaf{}
occupies about 40\% of the total gas mass, 
while this percentage can be up to about 80$\%$ 
or higher at high-$z$. 
Therefore at high-$z$ most of the gas lies on or around the $T$-$\Delta$ power-law relation.
Whereas the high-temperature low-density WHIM phase is negligible at high-$z$,
it appears significantly at low-$z$, 
resulting in puffy-looking gas distribution around the $T$-$\Delta$ power-law 
(see Fig. \ref{3fig:Nyx_Rho_T}), 
which makes $T$-$\Delta$ power-law fitting non-trivial at low-$z$.

We address this issue by implementing an improved fitting procedure 
following \citet{Villasenor2021}. 
First, we extract the temperature $T$ and the overdensity $\Delta$ for each cell 
of a simulation and then select gas with $-1.5 < \log \Delta < 0$ and $T < 10^5 K$ to avoid regions 
significantly deviant from the expected power-law $T$-$\Delta$ relationship.
Afterward, we divide the selected region into 15 equal-width bins in $\log \Delta$, where the overdensity $\log\Delta_i$ for each bin $i$ is given by the median value of overdensity in the bin. 
Here we define the bin temperature $\log {T_i}$ to be the maximum of the marginal temperature distribution $P (\log T\given \log \Delta_i)$ 
and its effective 1-$\sigma_{T,i}$ interval to be 1/2 of the temperature range containing the 68\% (16\% $\sim$ 84\%) 
highest probability density.  
The temperature-density relationship Eq.~(\ref{eqn:rho_T}) is
then fitted using a least squares linear fit 
on these $(\log\Delta_i, \log T_{\text{i}})$ pairs weighted by $ 1 / {\sigma_{\text{T,i}}}^2$. 
Examples of temperature-density relationships for two models in our thermal grid are illustrated in Fig.\ref{3fig:Nyx_Rho_T}. 
Power-law fits of the $T$-$\Delta$ relationship of our simulations are shown as white 
dashed lines while their values are given in the legends texts.
The peak temperature in each $\log \Delta$ bin  ($\log T_\text{peak,i}$,$\log\Delta_i$) are plotted as black 
dots and 1-$\sigma_{T,i}$ error bars are also shown. 
Left panel shows a model with $T_0=4353$ K and $\gamma=1.58$,
while right panel shows another model with $T_0=5091$ K and $\gamma=1.45$. 
Finally, as will be discussed later in \S\ref{sec:skewers}, 
we let the \HI{} photoionization rate $\Gamma_{\mathHI{}}$ be a free parameter when generating Ly$\alpha$ forest skewers from our simulations. 
As such, we add an additional parameter $\log \Gamma_{\mathHI{}}$ to our thermal grid,
extending it to [$\log T_0$, $\gamma$, $\log \Gamma_{\mathHI{}}$].
The value of $\Gamma_{\mathHI{}}$ we used in this study spans from
$\log (\Gamma_{\mathHI{}} /\text{s}^{-1})$ = -13.834 to $-12.932$
in logarithmic steps of $0.075$ dex, which gives 13 values in total (see right-hand panel of Fig.\ref{3fig:lowz_grid}). In total, the 3D thermal grid consists of $48\times 13=624$ models.

\subsection{Skewers}
\label{sec:skewers}

 \begin{figure*}
\centering
    \includegraphics[width=0.95\linewidth]{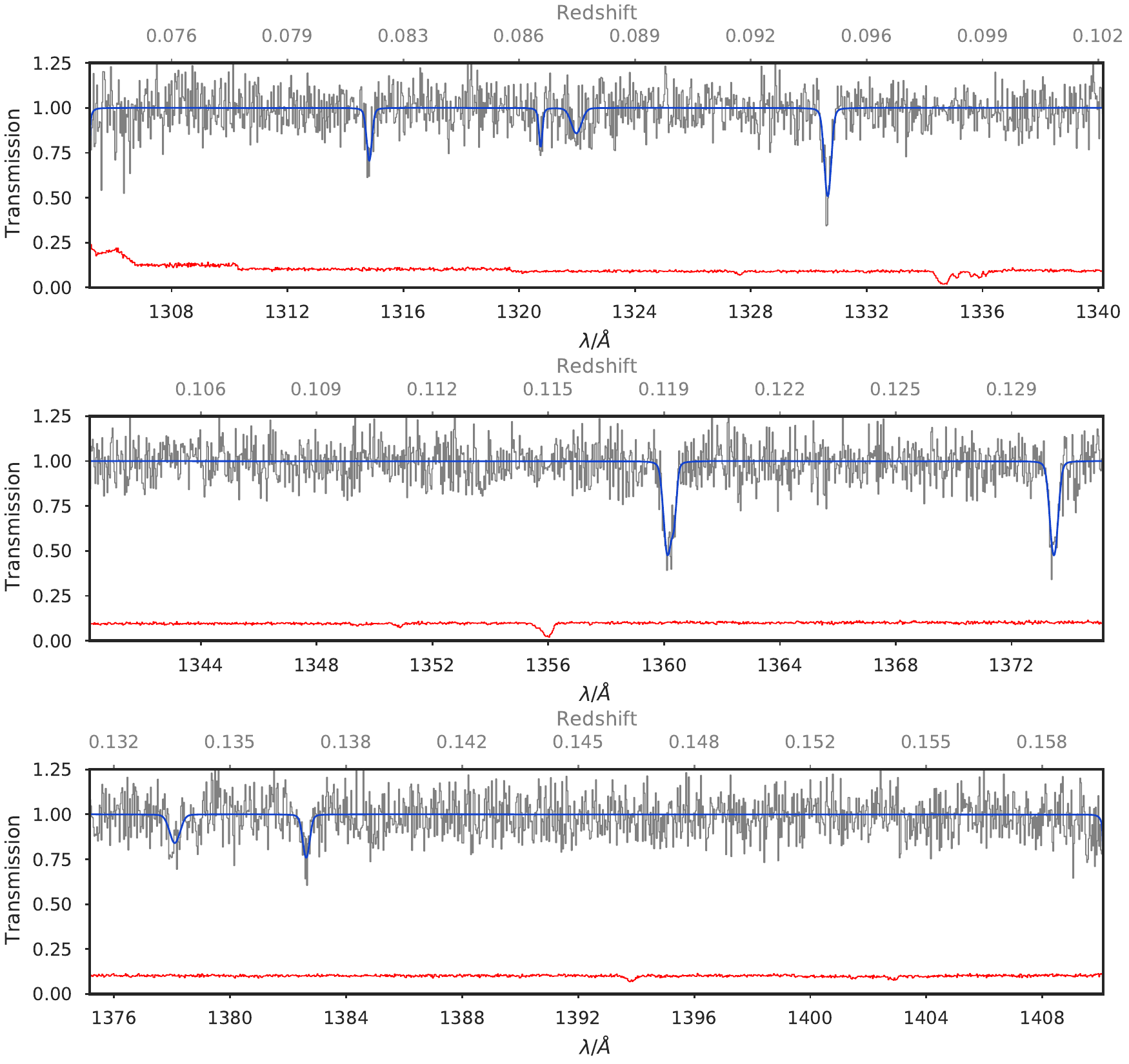}
  \caption{Illustration of one of our forward modeled spectra from Nyx simulation. The simulated raw spectrum is shown in gray, while a model spectrum based on \vpfit{} line fitting (described in \S~\ref{sec:vpfit}) is shown in blue and the noise vector is plotted in red. This particular spectrum is forward-modeled in order to model the instrumental effect and noise properties of one of the HST COS spectra in \protect\cite{Danforth2016} low redshift dataset.}
  \label{3fig:model_VP}
\end{figure*} 
We generate simulated \lya{}
spectra by calculating the Lyman-$\alpha$ optical depth ($\tau$) 
array along the line-of-sight, which hereafter will be referred as skewers for simplicity. 
For each model on the thermal grid, 
a set of 60,000 skewers are constructed parallel to the $x,y,z$
axes of the simulation box (20,000 skewers in each direction).
For each cell on these skewers, we extract properties needed for optical depth calculation, including temperature $T$, overdensity $\Delta$, and the velocity along the line-of-sight $v_z$. 
The hydrogen neutral fraction $x_{\mathHI{}}$, which is also needed to generate the synthetic Ly$\alpha$ forest skewers, 
is calculated by assuming ionization equilibrium while considering both collisional ionization and photoionization. 
Here the collisional ionization rate is computed based on the gas temperature $T$. 
Whereas the photoionization rate $\Gamma_{\mathHI{}}$ is set to be a free parameter in the post-processing of the simulation. 
Since Nyx does not model radiative transfer,
we approximate the self-shielding of the UV background 
for optically thick gas following the method given by \cite{Rahmati2013}, 
which amounts to attenuating  $\Gamma_{\mathHI{}}$ for cells containing dense gas. 

Given $x_{\mathHI{}}$, $T$, $\Delta$, $v_z$, and $\Gamma_{\mathHI{}}$,
we then calculate the optical depth $\tau$ in redshift space 
by summing contributions from all cells in real space along 
the line-of-sight following the full Voigt profile approximation by \citet{Tepper-G2006}. 
Then $F= e^{- \tau}$ gives us continuum normalized flux of \lyaf{} along with skewers. 
Lastly, we redo the procedure described above for each $\Gamma_{\mathHI{}}$ value to recalculate the skewers.
More specifically , we do not re-scale the $\tau$ to 
obtain skewers for a different $\Gamma_{\mathHI{}}$, which is the standard procedure at higher redshifts.  
This is because, whereas the high-$z$ \ac{IGM} is predominantly photoionized, 
there is significantly more shock-heated WHIM gas at low-$z$,
rendering the contribution from collisional ionization important
as shown by \citet{Khaire2019} for the case of \lya{} flux power spectrum. 
Although, it may not be essential for studying \lyaf~absorption lines,
to be more precise we recalculate skewers for each value of $\Gamma_{\mathHI{}}$.

\subsection{Forward Modeling of Noise and Resolution}
\label{sec:FM}
 
As discussed in \S\ref{sec:intro}, we are interested in understanding the constraints on 
the \ac{IGM} achievable with realistic data.
To this end, we generate mock datasets with properties consistent with the \citet{Danforth2016} low redshift Ly$\alpha$ forest dataset, which comprises 82 unique quasar spectra with ${\rm S\slash N} > 5$ observed with the \ac{COS} on the \ac{HST}. 
To avoid proximity regions and 
contamination from lower Ly-$\beta$, 
we use rest-frame wavelength range 
$1050-1180$ \AA{} 
to identify \lyaf{} for each of these spectra.
As a result, we select 34 of \citet{Danforth2016} quasar spectra covering the redshift range 
$0.06 < z < 0.16$ 
of our interest for the study, comprising a total redshift pathlength of 
$\Delta_{\text{ob}} = 2.136$,
which corresponds to our observational dataset for forward modelling.
We choose this redsfhit bin to be the same as the redshift bin used for power spectrum calculation by \citet{Khaire2019} at $z=0.1$ 
so that we can compare our future analysis with the results obtained with power spectrum measurements.

The \ac{COS} has a nominal resolution $R\sim 20000$, which corresponds to roughly $15$ km/s, 
and a non-Gaussian line spread function (LSF) exhibiting significantly broad Lorentzian wings, 
which could alter the shape of absorption lines on velocity 
scales larger than the resolution quoted above. 
For low-$z$ \ac{IGM} with temperatures at mean density $T_0\sim 5000~{\rm K}$,
the $b$-values for pure thermal broadening (i.e. the narrowest lines in the Ly$\alpha$ forest) are about $10 \sim 20$~{\rm km/s},
which means that the corresponding absorption features can not be fully resolved by \ac{COS}.
Thus, it is crucial to treat the instrumental effect carefully, 
including the peculiar shape of COS LSF. 
Therefore, we forward model noise and resolution to make our simulation results statistically comparable with the observation data.

In practice, we make use of tabulated COS LSF and noise vectors from \citet{Danforth2016} data.
For any individual quasar spectrum from the observation dataset,
we first stitch randomly selected simulated skewers without repetition to cover the same wavelength 
(in the rest frame $1050-1180$ \AA{}) of that quasar
and then rebin the skewers onto the pixels of the observed spectra. 
Then we convolve the simulated spectra with the HST \ac{COS} \ac{LSF} while taking into account the grating and life-time positions used for that specific data spectrum. Here the \ac{COS} \ac{LSF} is obtained from {\tt linetools}\footnote{For more information, visit https://linetools.readthedocs.io} and is tabulated for up to 160 pixels in each direction. We interpolate the \ac{LSF} onto the wavelengths of the mock spectrum (segment) to obtain a wavelength dependent \ac{LSF}. 
Each output pixel is then modeled as a convolution between the input stitched skewers and the interpolated \ac{LSF} for the corresponding wavelength. 
Afterward, the newly generated spectrum is interpolated to the wavelength of the selected \ac{COS} spectra. The noise vector of the quasar spectrum is propagated to our simulated spectrum pixel-by-pixel by sampling from a Gaussian with $\sigma = \psi_i$, with $\psi_i$ being the data noise
vector value at the 
i$^\text{th}$ pixel. In the end, a fixed floor 
is added to the error vector for all simulated spectra to avoid an artificial effect in post-processing,
which will be discussed later in \S\ref{sec:vpfit}.

For each model, 
we generated 2000 forward-modeled spectra, corresponding to a total pathlength $\Delta z_\text{tot} \sim 125$, from the 60,000 raw skewers\footnote{
For each Nyx model, 2000 spectra needs about 20,000 raw skewers, i.e, we randomly pick 20,000 skewers from 60,000.
}, 
and fit voigt profiles to each line in the spectra to obtain the \bn{} pairs for our dataset
(as described in section \S~\ref{sec:vpfit}).
For the purpose of illustration, an example of a forward-modeled spectrum is shown in Fig.\ref{3fig:model_VP} where the simulated spectrum is shown in gray, the model spectrum based on \vpfit{} line fitting (see \S~\ref{sec:vpfit}) is in blue, and the noise vector in red.

\subsection{VPFIT}
\label{sec:vpfit}

To perform the analysis based on the \bndist s, 
we have to fit the \lya{} lines in our simulated spectra to obtain a set of \bn{} pairs for each model.
To this end,
we run a line-fitting program on our forward-modeled mock spectra 
to obtain a set of $b$-$N_{\rm HI}$ pairs for each simulation model in our thermal grid. 
In this work, we use the line-fitting program \vpfit{}, 
which fits a collection of  Voigt profiles convolved with the instrument LSF to spectroscopic data \citep[][]{vpfit}\footnote{VPFIT: \url{http://www.ast.cam.ac.uk/~rfc/vpfit.html}}. 
Here, we employ a fully automated \vpfit{} wrapper adapted from \citet{Hiss2018},
which is built on the \vpfit{} version 10.2. The wrapper routine controls \vpfit{} with the help of the \vpfit{} front-end/back-end programs \texttt{RDGEN} and \texttt{AUTOVPIN} and fit our simulated spectra automatically.

 \vpfit{} identifies lines automatically and fits each line with three parameters:
the absorption redshift $z_{\text{abs}}$ of the line, its Doppler parameter $b$, and column density \NHI{}. \vpfit{} obtains these parameters for a collection of lines by minimizing the $\chi^2$ between the data and the model 
spectrum generated from all the fitted lines.
While fitting, \vpfit{} restrict $b$ and $N_{\rm HI}$ 
to $1 \leq b (\text{km/s}) \leq 300$ and $11.5 \leq \log (N_{\mathHI{}} / \text{cm}^{-2}) \leq 18$, respectively.
Our \vpfit{} wrapper allows us to fit spectra with a custom \ac{LSF}\footnote{Although 
our \vpfit{} wrapper allows us to implement an \ac{LSF} in \vpfit{}, only a single \ac{LSF} 
can be used at once, i.e. the wavelength dependence can not be 
taken into accout. 
As such, for the  input into \vpfit{} we use the \ac{LSF} at the lifetime of the data and 
evaluated it at the central wavelength of the spectrum that we are trying to fit. 
Such treatment is applied to both observed (mock) spectra and stimulated spectra so as to 
make sure our statistics are not biased.}. 
Since we are working at $0.06 \leq z \leq 0.16$, the Ly $\alpha$ forest lies completely in the wavelength range covered exclusively by the COS G130M grating having a central 
wavelength 1300 \AA{}. We fit our forward-modeled spectra with the same G130M LSF. 
Furthermore, the effective resolution of the grating also depends on the \ac{COS} lifetime
position during the observations, and they are also taken into account while running \vpfit~
as well as in forward modelling. An example of
model spectrum generated by combining lines fitted using \vpfit{} 
is shown in Fig.\ref{3fig:model_VP} as blue lines.

Moreover, we notice the presence of a significant number of absorption lines with very low Doppler-$b$ parameters and low column densities $N_{\mathHI{}}$ after fitting mock as well as real 
data with high signal-to-noise ratios (SNR). 
These weak narrow absorption lines, however, are not 
seen in our simulated and 
forward-modeled spectra. 
Visual inspection of these lines indicates that they are spurious and introduced
by \vpfit{} while attempting to fit artifacts due to flat-fielding, continuum placement, or errors in the data reduction. 
These lines are only introduced in 
spectra of the highest quality, where the extremely high \ac{SNR} leads to over-fitting by 
\vpfit{}. To avoid this problem, a fixed floor of value 0.02 
is added in quadrature to the error 
vector of the continuum normalized flux for all simulated spectra without 
adding additional noise to the normalized flux. With such a noise 'floor',
these weak features are essentially removed from the \vpfit{} output.
We find this floor value $0.02$ via trial and error. 
In practice, this additional noise floor mainly 
removes lines with $\log (\NHIt{}/\text{cm}^{-2})<12.5$ from
our dataset, which is outside our limits 
used in likelihood calculations 
(which will be discussed in 
\S\ref{3sec:log-likelihood})
and therefore not used in this study.

Furthermore, we follow the convention and 
apply another filter for both  $b$ and $ N_{ \mathHI}$ in this study,
using only $b$-$ N_{ \mathHI}$ pairs in region $12.5 \leq \log ( N_{\mathHI} / \text{cm}^{-2}) \leq 14.5$ and $0.5 \leq \log ( b / \text{km s}^{-1}) \leq 2.5$ in our analysis \citep{schaye2000,rudie2012,Hiss2018}. 
Such an limitation is chosen to include the \bndist s for all of our Nyx models 
while guaranteeing that the absorbers are not strongly saturated, which maximizes the
sensitivity to \ac{IGM} thermal state and minimizes the impact of poorly understood strong absorbers arising from the circumgalactic medium of galaxies.

 \section{Inference Algorithm}
  \label{sec:inference}
 
\citet{Hiss2019} introduced a Bayesian method to estimate the \ac{IGM} thermal parameters from the joint \bndist{}. 
In this paper, we adopt a similar approach while employing a new method for \bndist{} emulation, namely Density-Estimation Likelihood-Free Inference (DELFI). 
In addition, we also include the absorber number density along the line-of-sight \dndz{} in our analysis, 
i.e. the number of absorption lines (in some range of $b$ and $N_{\rm HI}$) per unit path-length along the line-of-sight, 
which helps us to better constrain the UV background photoionization rate $\Gamma_{\mathHI{}}$.
The reason behind this is that the \bndist{} is less sensitive to $\Gamma_{\mathHI{}}$ compared with thermal parameters $T_0$ and $\gamma$ (see Fig.\ref{fig:shiftpdf} and \S\ref{sec:parameter}),
whereas the number density of absorbers (see Fig.\ref{fig:emu_plot}) depends strongly on  $\Gamma_{\mathHI{}}$. 
It is analogous to the fact that the mean flux of the \lyaf{} is sensitive to $\Gamma_{\mathHI{}}$.
In this work, we emulate the \dndz{} using a Gaussian process emulator based on our simulations 
and employ it as a normalization factor in our likelihood function. 
More discussion about this modification is presented in \S\ref{3sec:log-likelihood} and Appendix~\ref{sec:2dresult}.

This section is organized as follows, we first introduce our new \bndist{} emulator
and then discuss the modifications to the likelihood function.
Afterward, we investigate the relationship between thermal parameters and \bndist{}  in \S\ref{sec:parameter}. Finally,  we present our inference results in \S\ref{sec:3dresult} and apply a series of inference tests to evaluate the statistical validity of our method in \S\ref{sec:Inference_test}.

 \subsection{Emulating the \bndist{} with DELFI}
 \label{sec:emu}

In this work, we build our \bndist{} emulator following the \ac{DELFI} method \citep{papamakarios2016, Alsing2018, papamakarios2018, Lueckmann2018, Alsing2019}, which turns inference into a density estimation task by learning the sampling distribution of the
data as a function of the parameters. 
Compared with the previously used \ac{KDE} method in \citet{Hiss2019}, this method provides a flexible framework for conditional density estimation and does not implicitly apply a smoothing kernel to the training data. It hence is able to deliver higher-fidelity conditional density estimators given the same training data. 

We make use of \texttt{pydelfi}\footnote{See https://github.com/justinalsing/pydelfi}
$-$ the publicly available \texttt{python} implementation of \ac{DELFI} based on \ac{NDE}s and
active learning \citep{Alsing2019}. \texttt{pydelfi} makes use of \ac{NDE}s to learn the sampling conditional probability distribution $P(\mathbf{d} \given \boldsymbol{\theta})$ of the data summaries $\mathbf{d}$, as a function of parameters $\boldsymbol\theta$, from a training set of simulated data summary-parameter pairs $\{\mathbf{d}, \boldsymbol\theta\}$.
In this work, the parameters $\boldsymbol{\theta}$ are $\log T_0$, $\gamma$ and $\log \Gamma_{\mathHI{}}$,
and the data summaries $\mathbf{d}$ are $\log \text{\NHI{}}$ 
and $\log b$\footnote{\texttt{pydelfi} also has the option to apply different data compression methods (e.g., \citealp{Alsing2018b}) and active learning methods to optimize the data and parameter space sampling. Here we do not exploit these features since we have pre-chosen our summary statistics and simulation grid (the \bndist{}) at a fixed grid of thermal parameters.},
and the \bndist\ is considered as a conditional probability distribution \bnpdfG{} learned from our simulations. 
More specifically, the \bndist{} is modeled as a Masked Autoregressive Flow (MAF; 
\citealp{Papamakarios2017}) neural density estimator, 
which is constructed as a stack of five Masked Autoencoders for Density Estimation, \citep[MADE;][]{Germain2015}, 
each with two hidden layers with $50$ units each and $\mathrm{tanh}$ activation functions.
The NDEs are trained by stochastic gradient descent. For more technical details about MAF and MADE neural network architectures see \citet{Germain2015}, 
\citet{Papamakarios2017} and \citet{Alsing2019}. To prevent over-fitting, the NDEs are weighted by their relative cross-validation losses and are trained with early-stopping (see \citealp{Alsing2019} for details).
For convenience, in this paper we will refer to the \bndist{} emulator discussed above as
the \ac{DELFI} emulator.

As mentioned above, the DELFI emulator is trained on 
the data summary-parameter pairs $\{ [\log T_0,\gamma, \log \Gamma_{\mathHI{}}] , [b, \log \text{\NHI{}}]\}$. 
For each model, we fit (VPFIT) 2000 simulated spectra,
 corresponding to a total pathlength $\Delta z_\text{tot} \sim 123$,
 to get \bn{} pairs for the model, 
and label these \bn{} pairs with their simulation parameters $[\log T_0,\gamma, \log \Gamma_{\mathHI{}}]$.  
Our training set therefore consists of all these labeled \bn{} pairs
for all models on the thermal grid. 
Here we quantify the size of data by its total pathlength 
rather than number of lines\footnote{ 
It means that the learned \bndist{} has a resolution that depends on the \dndz{} of the model.
We could instead set the number of \bn{} pairs to be fixed while using different total panthlength for each model.
However such a change does not affect the results of our inference method.
},
because the latter depends on the \dndz{} that varies among different models.

\subsection{Likelihood function}
  \label{3sec:log-likelihood}

   \begin{figure*}
 \centering
  \footnotesize
  \includegraphics[width=0.99\linewidth]{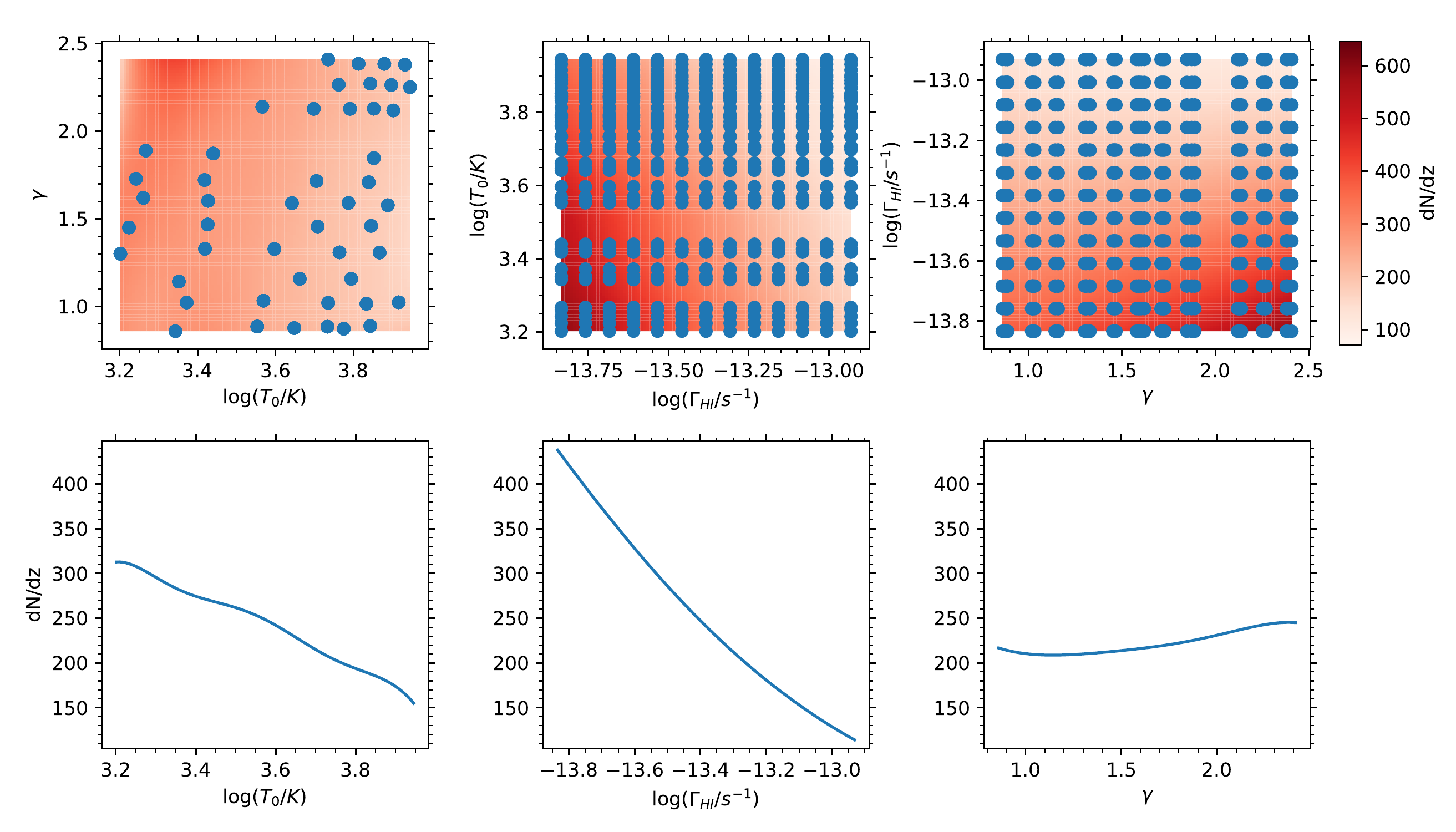}
  \caption{ An example of emulation of the absorber density \dndz{} generated by the Gaussian emulator sliced at the median value of the posterior from the MCMC process, where {$\log (T_0/\text{K})$} = 3.69, $\gamma=1.55$,  {$\log (\Gamma_{\mathHI{}} /\text{s}^{-1})$} = -13.30. Top panels are the 2D \dndz{} distributions where Nyx models are shown in blue circle. The top left panel is the \dndz{} on $\gamma$-$\log T_0$ plane at {$\log (\Gamma_{\mathHI{}} /\text{s}^{-1})$}= -13.30. The top middle is $\log T_0$-$\log \Gamma_{\mathHI{}}$ plane
  at $\gamma=1.55$. The top right is $\log \Gamma_{\mathHI{}}$-$\gamma$ plane at {$\log (T_0/\text{K})$} = 3.69. 
  Bottom panels are marginalized 1D \dndz{} distributions at the thermal parameters mentioned above.
  From left to right:  \dndz{} vs $\log T_0$, \dndz{} vs $\log \Gamma_{\mathHI{}}$, and \dndz{} vs $\gamma$.
  }
  \label{fig:emu_plot}
\end{figure*}

\citet{Hiss2019} used only the shape of \bndist{} to constrain IGM thermal 
parameters, but ignored the normalization, which can be thought of as the total number of absorption lines in the dataset or equivalently as the line density $dN\slash dz$. 
Here we generalize the likelihood formalism introduced in \citet{Hiss2019} 
to include the information contained in the absorber density \dndz{} \citep[see also][]{hissthesis}. 
Our goal is to find the likelihood of observing a set of absorption lines $\left\{b_i, N_{\mathHI{},i}\right\}$ given a model with a set of thermal parameters [${\log T_0}^\prime, {\gamma}^\prime , {\log \Gamma_{\mathHI{}}}^\prime$].
We first assume that the \ac{PDF}s are normalized such that
\begin{equation}
\iint{P(b, { N_{\mathHI{}} )\, \text{d}{ N_{\mathHI{}}} \,\text{d}b}} =1,
\label{eq:integral}
\end{equation}
where $P(b, { N_{\mathHI{}}})$ is the conditional probability distribution function 
${P(b \mathbin{,} N_{\mathHI{}}\,\given\,{T_0}^\prime, {\gamma}^\prime,{\Gamma_{\mathHI{}}}^\prime)}$,
for simplicity we write it as $P(b, { N_{\mathHI{}}})$ in the rest of this subsection.
We imagine dividing the $b$-$N_{\rm HI}$ into a set of infinitesimally fine grid cells, 
such that the occupation number of each grid cell is either one or zero.
Knowing that our set of observational/mock dataset $\left\{b_i, N_{\mathHI{},i}\right\}$ is comprised of $n$ lines,
and assuming that there are $N_\text{g}$ grid cells in total, 
the likelihood for a model with thermal parameters [${\log T_0}^\prime, {\gamma}^\prime , {\log \Gamma_{\mathHI{}}}^\prime$]
can thus be written as the following product of Poisson probabilities\footnote{In assuming the probability distribution 
for each grid cell is Poisson, we are implicitly assuming each $b$-$N_{\rm HI}$ pair is an uncorrelated draw from the \bndist{}. This assumption,  also made by \citet{Hiss2019}, 
amounts to ignoring the spatial correlations between absorption lines.  \citet{Hiss2019}
showed that this is a very good approximation and yields unbiased inference as we will also demonstrate in \S~\ref{sec:Inference_test})}
\begin{align}\label{eq:likelihood1}
\mathcal{L}= &P(\mathrm{data}|\mathrm{model})\\ \nonumber
=&\left( \prod_{i=1}^{n} \mu_{i}\,e^{-\mu_{i}} \right) \left( \prod_{j\ne i}^{N_\text{g}} e^{-\mu_{j}} \right), 
\end{align}
where the first product is over the occupied cells, and the second product is over the empty
cells. 
Here the $\mu_i$ is the Poisson rate of occupying a cell in the $b$-$N_{\rm HI}$ plane with area $\Delta {\rm N_{\mathHI{}}}_i\times \Delta b_i$, i.e.
\begin{equation}
\mu_{i}=\left(\frac{\text{d} N}{\text{d} z}\right)_{\rm model}\,P(b_i, N_{\mathHI{},i})\,\Delta { N_{\mathHI{}}}\, \Delta b\, \Delta z _{\rm data}, 
\label{eq:mu}
\end{equation}
where $P(b_i, { N_{\mathHI{}}}_i)$ 
is the probability distribution function evaluated at the point $(b_i, N_{\mathHI{},i})$ 
using the DELFI \bndist{} emulator described in \S~\ref{sec:emu}, 
and $\Delta z_{\rm data}$ is the total redshift path covered by the data spectra from which we obtain our data set \bn{},
whereas $\left({\text{d} N}\slash{\text{d} z}\right)_{\rm model}$ is the absorber density of the model which will be further discussed later in this subsection. 

Afterwards, it is easy to show that Eq.~(\ref{eq:likelihood1}) implies 
\begin{equation}\label{eq:log-likelihood}
\ln \mathcal{L} = \sum_{i=1}^{n} \ln (\mu_i)  - \sum_{k=1}^{N_g} \mu_{k}.
\end{equation}
Above, the second sum over $k$ is simply an integral of Eq.~(\ref{eq:mu}) over the $b$-$N_{\mathHI{}}$ plane,
while the integral of 
$P(b, { N_{ \mathHI}})$ $\text{d}{ N_{ \mathHI} } \,\text{d}b$ over the plane is unity according to Eq.~(\ref{eq:integral}). 
As a result, we can write our likelihood function as
\begin{equation}
\ln \mathcal{L} = \sum_{i=1}^{n} \ln (\mu_i) - \left(\frac{\text{d} N}{\text{d} z}\right)_{\rm model}\Delta z _{\rm data}. 
\label{eq:likelihood}
\end{equation}

Since \citet{Hiss2019} did not consider the absorber density,
the likelihood in their analysis is simply given by $\ln \mathcal{L_\text{Hiss}} = \sum_{i=1}^{n} \ln P(b_i, N_{\mathHI{},i})$.
In comparison, our likelihood function Eq.~(\ref{eq:likelihood}) can be written as 
\begin{equation}
\label{eq:likelihood2}
\ln \mathcal{L} = \sum_{i=1}^{n} \ln P(b_i, N_{\mathHI{},i}) + n\ln \xi - \xi,
\end{equation}
where $\xi = ( {\text{d} N } /{\text{d} z} )_{\rm model} \Delta z _{\rm data}$.
We can see that the first term remains the same, 
and our modification (the implementation of absorber density \dndz) can be considered as a correction term based on the absorber density of the model,
the number of lines observed, and the pathlength of the data set $\Delta z_\text{data}$.

As a result of our modification, the likelihood of observing a line with certain line parameter $(b, N_{\mathHI{}})$ now depends not only on the \bndist{}s of models but also on absorber densities of the models. 
Consequently, to evaluate the likelihood $\mathcal{L}$ on the parameter space, 
we need the ability to evaluate $\left({\text{d} N}\slash{\text{d} z}\right)_{\rm model}$ at an arbitrary location on the parameter space. 
To this end, a Gaussian process emulator (based on \texttt{George}, see \citealt{Ambikasaran2016}) is employed to
emulate $\left({\text{d} N}\slash{\text{d} z}\right)_{\rm model}$ by interpolating the \dndz{} of models from Nyx simulations
based on their $N_{\text{model}}/\Delta z_{\text{model}}$,
where $\Delta z_{\text{model}}$ is the total pathlength of simulated spectra that are fed into \vpfit{}, 
and $N_{\text{model}}$ is the total number of lines identified by \vpfit{} from these spectra. 
The Gaussian process emulator is constructed with smoothing lengths of 40\% of our thermal grid length\footnote{ The smoothing length is input as initial guess, which is then refined later in the routine. In addition, all dimensions in the thermal grid are rescaled to unity in the Gaussian process emulator.} 
in $\log T_0$ and $\log \Gamma_{\mathHI{}}$ and a smoothing length of 80\% of thermal grid length in $\gamma$. 
The longer smoothing length in $\gamma$ is set to prevent the emulator from over-fitting the noise, 
considering that $\gamma$ has less effect on the absorber density \dndz{} compared with $T_0$ and $\Gamma_{\mathHI{}}$ (see Fig.\ref{fig:emu_plot}), 
which makes small fluctuations induced by noise more significant.

The results of our \dndz{} emulation are shown in Fig.\ref{fig:emu_plot},
where both $\log T_0$ and $\log \Gamma_{\mathHI{}}$ (left and middle column) have negative correlations with absorber density \dndz{}. 
This dependence  can be explained qualitatively by the fluctuating Gunn-Peterson approximation \citep[FGPA, see][]{Weinberg1997}
\begin{equation}
\tau_{\text{Ly} \alpha} \propto n_{\rm HI} \propto x_{\rm HI} n_{\rm H} \propto \frac{n^2_{\rm  H} T^{-0.7}}{\Gamma_{\rm HI}},
\label{eq:tau}
\end{equation}
where the $\tau_{\text{Ly} \alpha}$ denotes the \lya{} optical depth and the $n_{\rm H}$ is the hydrogen number density. 
This equation implies that both higher temperatures and higher photoionization rates reduce the \lya{} optical depth of gas absorbers in the \ac{IGM},
leading to lower absorber density.
The wiggles shown in \dndz{} vs $T_0$ plot (bottom left panel of Fig.\ref{fig:emu_plot}) are effects of poor interpolation due to lack of models at $\gamma \sim 1.5 $ (see top left panel). 
Moreover, we notice a weak correlation between $\gamma$ and \dndz{} (see the {bottom right} panel of Fig.\ref{fig:emu_plot}).
{However, such $\gamma$ dependence is relatively weak compared with $T_0$ and $\Gamma_{\rm HI}$ dependencies,
and is likely caused by artifacts due to the emulation. As shown in the top left-hand panel, we do not have models in low $T_0$ high $\gamma$ region, 
the absorber density \dndz{} could thus be 
over-extrapolated in these regions, 
further biasing the $\gamma$ dependence on the whole parameter space. 
We performed some tests and found that the weak correlations in 
$\gamma$ - \dndz{} vanishes if we do not include the high $\gamma$
simulations. 
Therefore, in conclusion, the marginalized $\gamma$ - \dndz{} correlation shown in Fig.~\ref{fig:emu_plot} is an artifact introduced by our Gaussian
emulator, however, it is too weak to affect our inference results.}

\subsection{Parameter study}
\label{sec:parameter}

\begin{figure*}
 \centering
    \includegraphics[width=0.99\linewidth,keepaspectratio]{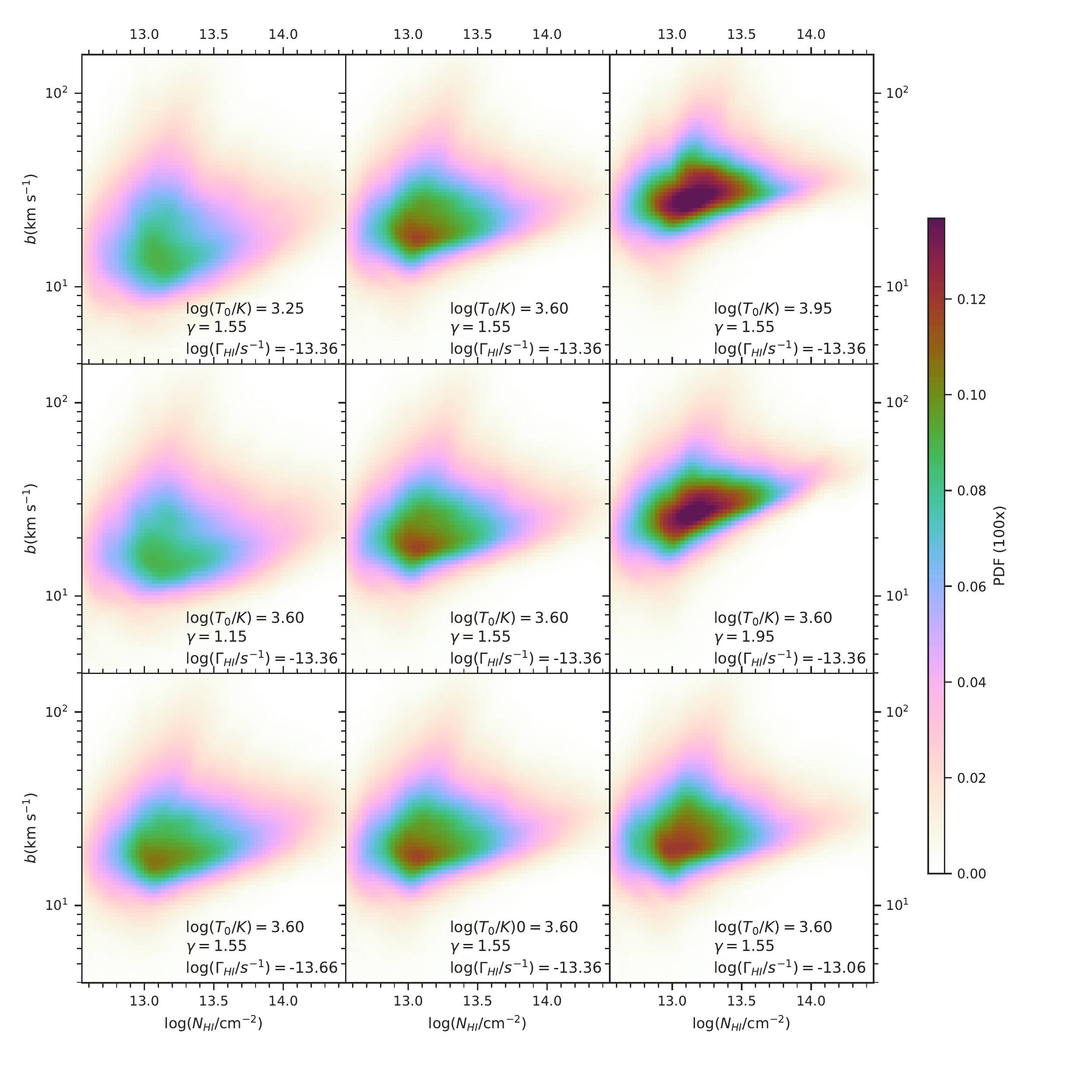}
    \vspace{-0.9cm}
  \caption{Comparisons of \bndist s modeled by \ac{DELFI} emulator with different thermal parameters. 
  Top panel shows changes in the \bndist{}  with increasing $\log T_0$, where {$\log (T_0/\text{K})$} = 3.25 (left), 3.60 (middle) and 3.95 (right) respectively, while $\gamma$=1.55 
  and {$\log  (\Gamma_{\mathHI{}} /\text{s}^{-1})$}=-13.36 for all three plots. The middle panel shows changes of the \bndist{} where $\gamma$ =1.15 (left), 1.55 (middle) and 1.95 (right) respectively, while {$\log (T_0/\text{K})$}=3.60 
  and {$\log  (\Gamma_{\mathHI{}} /\text{s}^{-1})$}=-13.36 are fixed. The bottom panel shows \bndist s with decreasing UV background. {$\log  (\Gamma_{\mathHI{}} /\text{s}^{-1})$}= -13.66 (left), -13.36 (middle) and -13.06 (right), 
  while $\log T_0$ and $\gamma$ remain unchanged. 
  All pdfs here are normalized to unity. 
  For illustration purposes, values of pdf are multiplied by 100 in the color bar.}
  \label{fig:shiftpdf}
\end{figure*}

A new feature of the DELFI \bndist{} emulator is its ability to emulate \bndist s continuously on the parameter space. 
With such a feature, 
we are now able to illustrate the parameter dependence of the \bndist{} 
and investigate the physics behind these dependence.
Fig.\ref{fig:shiftpdf} shows emulated \bndist s with different values of thermal parameters [$\log T_0$, $\gamma$ , $\log \Gamma_{\mathHI{}}$].
The top panel shows \bndist s with increasing $T_0$, where {$\log (T_0/\text{K})$} = 3.25 (left), 3.60 (middle) and 3.95 (right) respectively, 
while $\gamma$=1.55 and $\log (\Gamma_{\mathHI{}} /\text{s}^{-1}) $=-13.36 for all three plots. 
Increasing $T_0$ results in the upward shifting of the \bndist s, 
which can be explained by the thermal component of the $b$ parameter and the  $T$-$\Delta$ relationship Eq.~(\ref{eqn:rho_T}), i.e.
\begin{equation}
b_T \propto (2kT/m)^{1/2} \propto (T_0 \Delta^{\gamma -1})^{1/2},
\label{eq:b_T}
\end{equation}
where higher $T_0$ results in higher \ac{IGM} temperature, leading to larger $b$ parameters. In addition, we notice that as the $T_0$ goes up,
the \bndist{} becomes more concentrated, i.e. the distribution becomes tighter, and the pdf values increases. 
Such behavior might be explained as follows. 
There are two components contributing to $b$ parameter, 
namely thermal motion and {non-thermal broadening}.  
The thermal component is associated with the \ac{IGM} temperature
and thus follows a distribution determined by $T_0$.
On the other hand, as a result of the small-scale motion of the gas,
the {non-thermal} component is independent of the temperature and has a large dispersion, leading to broader distribution. 
At low temperatures, where the thermal contribution is weak, 
the $b$ parameter is dominated by the {non-thermal} component, 
resulting in broad distribution. 
As the temperature goes up, 
the thermal component dominates over {non-thermal broadening},
and the $b$ parameter thus concentrates on a central value of $b$ 
determined by the IGM temperature.

The middle panel of Fig.\ref{fig:shiftpdf} shows the \bndist{} with increasing $\gamma$, 
where $\gamma$ =1.15 (left), 1.55 (middle), and 1.95 (right), respectively, 
while {$\log (T_0/\text{K})$} =3.65 and $\log  (\Gamma_{\mathHI{}} /\text{s}^{-1})$=-13.36 are fixed. 
These plots indicate that there are degeneracies between $\gamma$ and $T_0$, 
where an increasing $\gamma$ also shifts \bndist{}s upwards,
which can be understood from Eq.~(\ref{eq:b_T}) 
and the fact that 
at low-$z$, the \lya{} lines originate predominantly from gas with $\Delta_\text{abs} > 1$ 
($\Delta_\text{abs} \sim 10$, see \citealt[][]{Gaikwad2017}),
which results in higher temperatures at densities of absorbers for models with larger $\gamma$.
The concentration effect is also seen in the middle panel,
which can be explained in the same way as the upward shifting of the \bndist{} due to increasing~$\gamma$. 
It can also be seen from the middle panel that the $\gamma$ is correlated with the slope of the low-$b$ cutoff of the \bndist, 
which is consistent with the analytical fit of the low-$b$ cutoff,
where the slope can be approximated by $\Delta\log b/\Delta \log N  = (\gamma -1)/3$ \citep[see][]{rudie2012}.

The bottom panel of Fig.\ref{fig:shiftpdf} shows \bndist s with increasing photoionization rate $\Gamma_{\mathHI{}}$, 
where $\log  (\Gamma_{\mathHI{}} /\text{s}^{-1})$= -13.66 (left), -13.36 (middle) and -13.06 (right), 
while $\log T_0$ and $\gamma$ remain unchanged. 
We observe that increasing $\Gamma_{\mathHI{}}$ results in a similar but much weaker effect compared with increasing $T_0$, i.e. the \bndist{} slightly shifts upward and becomes more concentrated with increasing $\Gamma_{\mathHI{}}$.
Such effects are because the photoionization rate $\Gamma_{\mathHI{}}$  alters the \lya{} optical depth of the \ac{IGM}.
Since the \lyaf{} typically probes regions with optical depth $\tau_{\text{Ly} \alpha} \sim 1$, given higher $\Gamma_{\mathHI{}}$, 
it probes regions with higher temperatures and densities, 
which can be derived from Eq.(\ref{eq:tau}), 
causing effects similar to increasing $T_0$.
However, such effects are relatively weak, making the \bndist{} less sensitive to the photoionization rate $\Gamma_{\mathHI{}}$. 

All these aforementioned parameter dependences (except $\Gamma_{\mathHI{}}$, which is not considered in previous works ) of the \bndist{} are consistent with previous works that measure the \ac{IGM} thermal state based on the full \bndist{}\footnote{
In \citet{Hiss2019}, at $z \sim 2$, the \lya{} lines originate predominantly from gas with $\Delta < 0$, causing different effects when changing $\gamma$. However, the physics explanations behind the effect are coherent.
} 
\citep{Hiss2019} and low-$b$ cutoff \citep[][]{schaye1999,rudie2012,bolton2014,Rorai2018,Hiss2018},
indicating that our DELFI emulator successfully reproduce the parameter dependences of the \bndist{}. Furthermore, it also implies that our understanding of the \bndist{} agrees with the physics prediction.

\subsection{Inference results}
\label{sec:3dresult}

 \begin{figure*}
 \centering
    \includegraphics[width=0.85\linewidth]{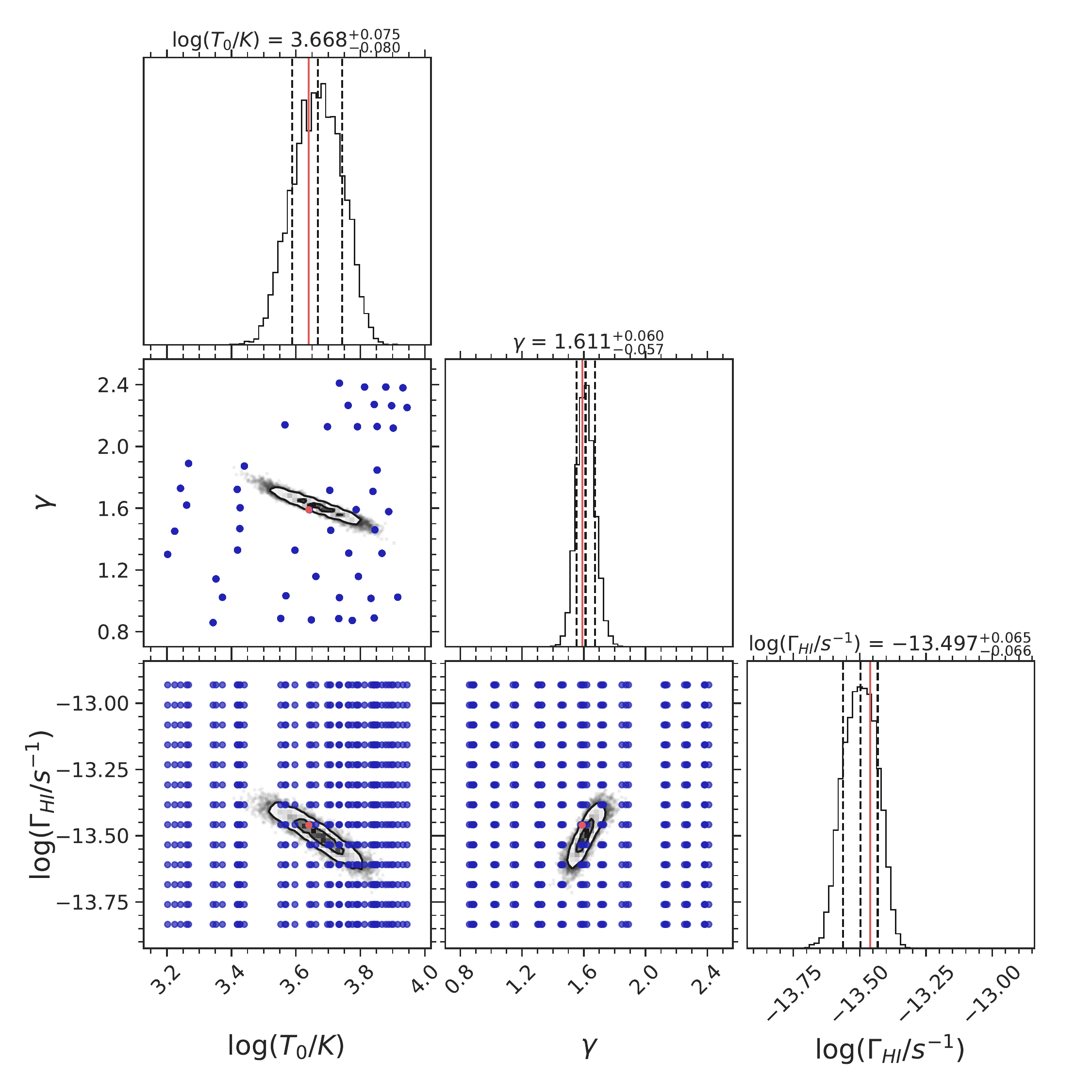}
  \caption{MCMC posterior for one of the models from Nyx simulation (absorbers shown in Fig.\ref{fig:fit_Nyx}) using the likelihood function~Eq.~\ref{eq:likelihood}. Projections of the thermal grid used for generating models are shown as blue dots, while the true model is shown as red dot. Inner (outer) black contour represents the projected 2D 1(2)-sigma interval. The parameters of true model 
  are indicated by red lines in the marginal distributions, while the dashed black lines indicates the 16, 50, and 84 percentile values of the posterior. The true parameters are: {$\log (T_0/\text{K})$} = 3.643, $\gamma=1.591$ and {$\log (\Gamma_{\mathHI{}} /\text{s}^{-1})$} = -13.458.  }
  \label{fig:corner_Nyx}
\end{figure*}

 \begin{figure}
 \centering
    \includegraphics[width=\columnwidth]{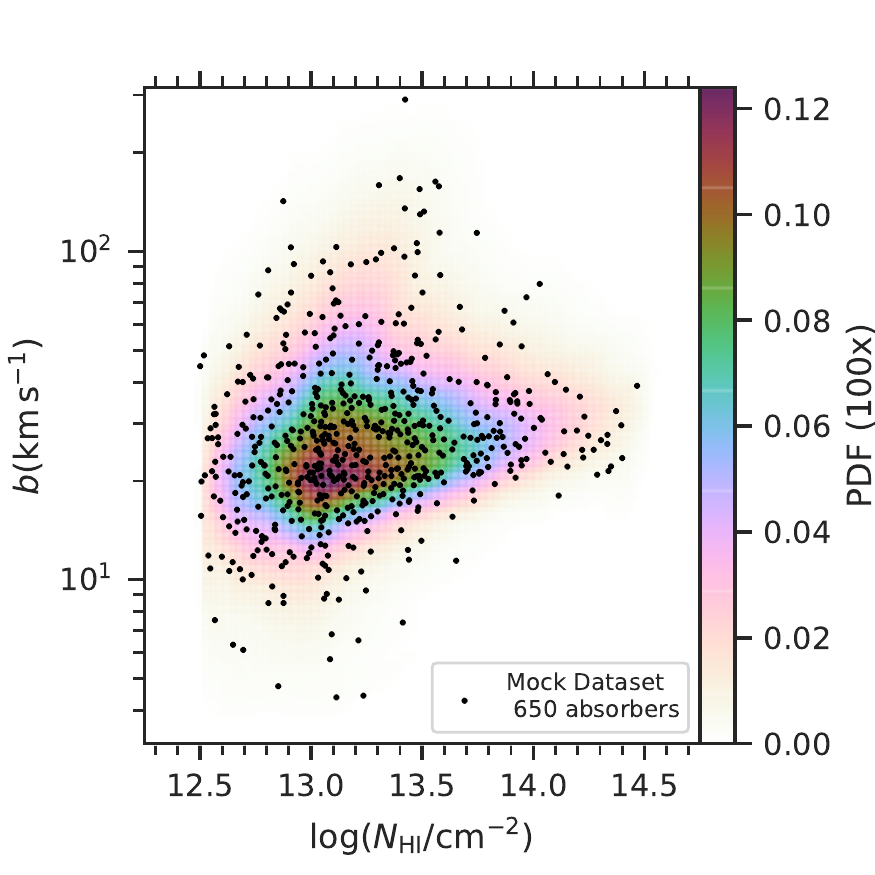}
  \caption{The color map is the full \bndist{} recovered from the Nyx mock dataset, which is emulated by our DELFI emulator based on the best-fit parameters (median values of the marginalized MCMC posterior), where {$\log (T_0/\text{K})$} = 3.668, $\gamma=1.611$ and {$\log (\Gamma_{\mathHI{}} /\text{s}^{-1})$} =-13.498. Black dots are the mock datasets we used in the inference. For illustration purposes, values of pdf are multiplied by 100 in the color bar.}
  \label{fig:fit_Nyx}
\end{figure}

 \begin{figure*}
 \centering
    \includegraphics[width=0.85\linewidth]{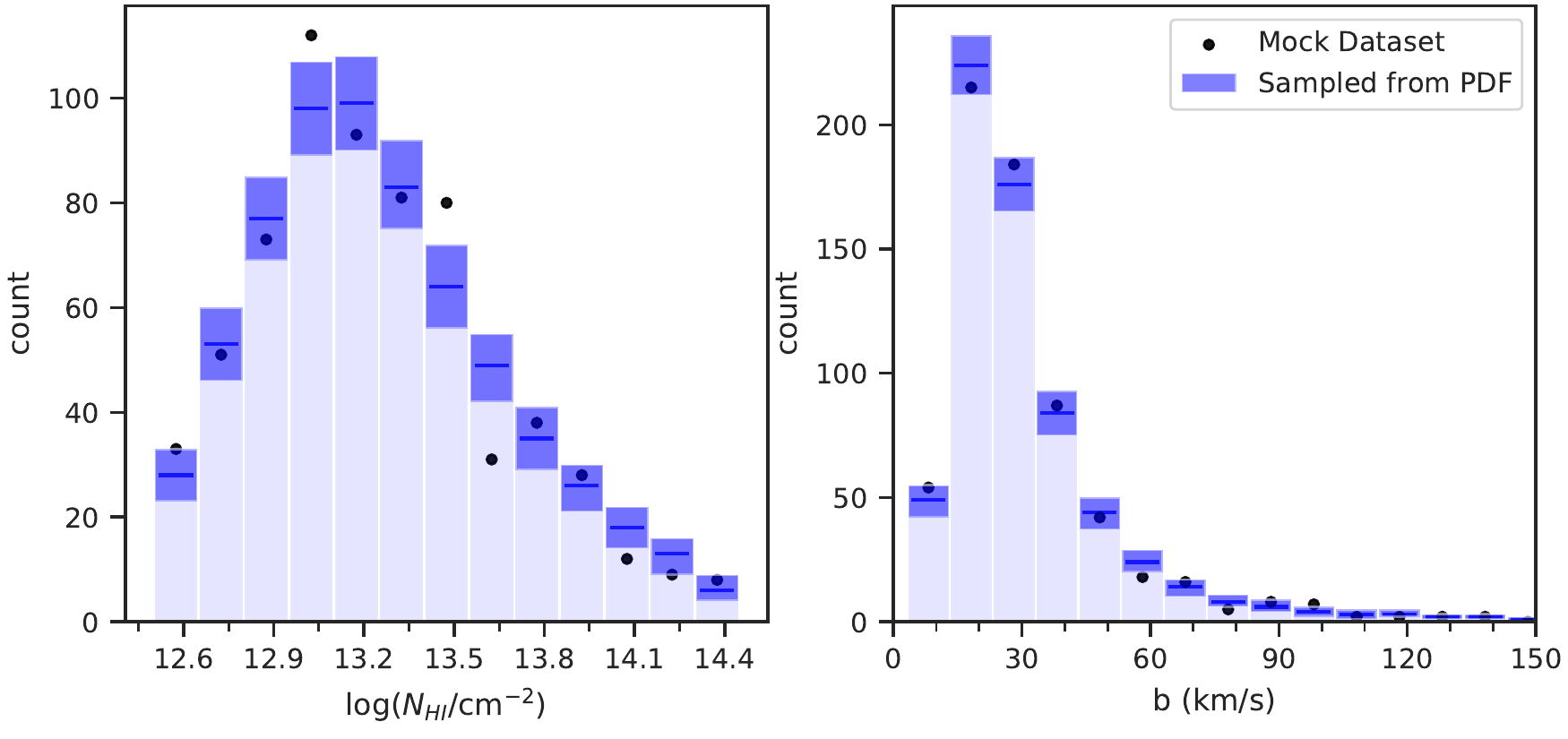}
  \caption{Marginalized 1D distributions of $N_{\mathHI{}}$ (\emph{left-hand panel}) and   $b$  (\emph{right-hand panel}) for the mock dataset (black dots) and the sampling from emulated \bndist~(blue bars) at {$\log (T_0/\text{K})$} = 3.699, $\gamma=1.549$ and {$\log (\Gamma_{\mathHI{}} /\text{s}^{-1})$} =-13.506. 
  Blue bars show the average of 5000 sampling from the emulated \bndist~using MCMC, while the (dark) blue shaded regions represent the 1-$\sigma$ fluctuation (16\%-84\% percentile among 5000 samples).
  }
  \label{fig:bN_1dhist}
\end{figure*}

Sets of mock spectra are created from our Nyx simulations to test the performance of our inference algorithm under realistic conditions.  
These mock spectra set are generated from a 
set of simulated spectra following a forward-modeling 
approach designed to match the pathlength, resolution, 
and noise properties of the \citet{Danforth2016} low-redshift quasar spectra in one-to-one correspondence as described in \S\ref{sec:FM}. 
Consequently, each mock spectra set consists of 34 forward-modeled spectra, which has exactly the same noise vectors, instrumental effects, and total pathlength ($\Delta z_{\text {data}}$=2.136) as the real observed dataset,  
which ensures that the accuracy of our analysis is realistic and achievable when the method is applied to real data.
A set of $\left\{b,N_{\rm HI}\right\}$ pairs, obtained by fitting these spectra using \vpfit{} (see  \S\ref{sec:vpfit}), 
is then used as the 'data' in the likelihood function (see Eq.\ref{eq:likelihood1}) to infer the posterior
distribution for IGM thermal parameters for this mock dataset. 

In this work, we perform inference via \ac{MCMC} sampling using the \texttt{python} package \texttt{emcee} \citep{foreman-mackey2013emceeMCMChammer}, 
which implements the affine-invariant sampling technique \citep{goodman2010Ensemblesamplersaffine} to sample the posterior probability distribution. 
Here the posterior is calculated based on the likelihood in Eq.~(\ref{eq:log-likelihood}),
which takes into account the absorber density \dndz{} as described in \S\ref{3sec:log-likelihood}, 
while assuming uniform (flat) priors for $\log T_0$, $\gamma$ and $\log \Gamma_{\mathHI{}}$, 
where the boundaries are chosen to be the range of each respective parameter in 1D. 
\ac{MCMC} posteriors obtained from the aforementioned mock datasets ($\left\{b,N_{\rm HI}\right\}$ pairs) are shown in Fig.\ref{fig:corner_Nyx}. 
We obtain $\log (T_0/\text{K}) = 3.668^{+0.075}_{-0.080}$, $\gamma=1.611^{+0.060}_{-0.055}$ and $\log (\Gamma_{\mathHI{}} /\text{s}^{-1}) = -13.497^{+0.065}_{-0.066}$ from the marginalized distributions, 
whereas the true parameters are: $\log (T_0/\text{K})  = 3.643$, $\gamma=1.591$ and $\log (\Gamma_{\mathHI{}} /\text{s}^{-1}) = -13.458$ (red dot and red vertical lines).
We recover the input parameters in very high precision with errors $\Delta \log (T_0/\text{K})  = +0.025$dex, $\Delta  \gamma = +1.3\%$, 
and $\Delta  \log (\Gamma_{\mathHI{}} /\text{s}^{-1})= -0.039 $dex,
while true parameters (red dot/solid lines) are all in the 1-$\sigma$ interval (inner black contours/ black dashed lines) of the posterior.
Here the degeneracy between $T_0$ and $\gamma$ can be quantitatively understood by the $T$-$\Delta$ relationship Eq.~(\ref{eqn:rho_T}) and the typical overdensity of absorbers $\Delta_\text{abs} \sim 10$. 
More specifically, both higher $T_0$ and $\gamma$ result in higher temperature of the absorbers, shifting the \bndist{} upward (see Fig.\ref{fig:shiftpdf} and relevant discussion in \S\ref{sec:parameter}). 
The degeneracy between $T_0$ and $\Gamma_{\mathHI{}}$ is mainly a result of the degeneracy in 
the absorber density \dndz{} with respect to the two parameters (see Fig.\ref{fig:emu_plot} and Fig.\ref{fig:corner_Nyx_NOdndz} as comparison), which is explained in ~\S\ref{3sec:log-likelihood}. 
It is noteworthy that our inference algorithm provides preeminent accuracy
for all three parameters even under a very realistic condition, 
where the resolution of spectra is rather low (with lines not fully solved), 
and the number of data is limited (with a total pathlength $\Delta_z=2.136$). 
Such a high sensitivity and precision makes our inference method a 
powerful tool in the study of the low-$z$ \ac{IGM} and \lyaf{}.

Fig.\ref{fig:fit_Nyx} shows the full \bndist{} recovered from the mock dataset,
which is emulated by our DELFI emulator based on the best-fit parameters (median values of the marginalized MCMC posterior). 
It appears that the PDF (color map) successfully represents the density distribution of the data points.
Furthermore, marginalized 1D distributions of $b$ and $N_{\mathHI{}}$ are given in Fig.\ref{fig:bN_1dhist} for both the mock dataset (black dots) and random samples from the emulated \bndist{} (blue bars). 
It can be seen that our emulator successfully reproduces the 1D marginalized $b$ and $N_{\mathHI{}}$ distribution, 
though there is a fluctuations in $N_{\mathHI{}}$ for the mock dataset at around $\log (N_{\mathHI{}}/{\text{cm}^{-2}}) \sim 13.5$. 
We figured out that such fluctuation is caused by the random error during the generation of the mock dataset,   
which can be reduced by increasing the size of the mock datasets.
However, to test the performance of our inference method under realistic conditions, 
we fix the size of the mock datasets and bear with such fluctuation in this work.

\subsection{Inference test}
 	\label{sec:Inference_test}

As discussed above, the likelihood function used in our inference algorithm involves
several approximations and emulation/interpolation procedures.
Most importantly, our inference ignores correlations between the lines \citep[see the discussion in][]{Hiss2019},
and we emulate the \bndist{} and the \dndz{} with our DELFI and Gaussian emulators respectively,
while both emulations involve interpolations.
These procedures might induce additional uncertainties
that are counted in our error budget\footnote{
The uncertainty of the \bndist s emulated by \ac{DELFI} is also ignored in our analysis. 
Such uncertainty is caused by the randomness in the training process, 
and has not been included in the results. 
But since our inference method (and the toy model) does well in the inference test, 
such randomness should be smaller than stochastic error shown in our analysis, 
and should not dominate our error budget. 
},
we hence want to make sure our inference results are valid under 
these assumptions, and our interpolation procedures work correctly.
Therefore, we perform a series of inference tests
to evaluate the robustness of the entire inference method.
An inference test is to carry out
a set of realizations of the inference algorithm based on the mock dataset 
and inspect the results to reveal 
if the inference method returns valid posterior probability distributions, 
i.e. whether the 'true model' is included in a set of probability contours following the ratio indicated by the posterior. 

The inference test is done as follows. 
First of all, we adopt the same prior as described in \ref{sec:3dresult}, 
and construct a regular uniform grid in the parameter space spanning the range set by our prior.
For each realization, we pick a model (set of parameters) on the above grid,
which we refer to as the `true model'.
and we refer its thermal parameters as 'true parameters' $\pmb{\theta}_{\text{true}}$,
We then create a corresponding mock dataset following the prescription described in \S\ref{sec:3dresult}. 
Given the mock dataset, since our priors are flat, 
we can determine the corresponding posterior probability distribution 
by evaluating the likelihood function $\mathcal{L}=P(\mathrm{data}|\mathrm{model})$ on the whole parameter space.
We then normalize the posterior function to unity and
determine 3D posterior probability contours based on the posterior (likelihood) distribution.
Knowing that the likelihood function is continuous on the whole domain, 
the 3D volume integral can hence be substituted by a 1D integral over the sorted likelihood function.
Here we define the probability contours $C_P$ and the likelihood thresholds $\mathcal{L}_{P}$ in the following way,
\begin{equation}
\iiint_{C_{P}} \mathcal{L} \text{d}V = \int_{{\mathcal{L}_{P}}}^\infty \mathcal{L}  \text{d}\mathcal{L}=P,
\label{eq:L_P}
\end{equation}
such that a probability contour $C_P$ is simply where $\mathcal{L}=\mathcal{L_\text{P}}$,
and any 'model' with parameter $\pmb{\theta}$ being inside a contour $C_P$ 
thus becomes equivalent to $\mathcal{L(\pmb{\theta})}  > \mathcal{L_\text{P}} $.
We further define the effective 1$\sigma$ (68\%) and 2$\sigma$ (95\%) intervals 
as the volume between contour pairs $(C_{0.16},C_{0.84})$ and $(C_{0.025},C_{0.975})$ respectively.
Finally, we judge the performance of our inference method based on 
how often the parameters of the 'true model'
$\pmb{\theta}_{\text{true}}$ falls 
in these 1(2)-$\sigma$ interval contour pairs compared to the expectation based on the corresponding probabilities, i.e.
if our posterior distribution is perfect, the true model 
should land within the  1$\sigma$ (2$\sigma$) contours 68\% (95\%) of the time.
An example of the distribution of the likelihood function is shown in Fig.\ref{fig:inftest_1}, and more details about the calculation of the likelihood distribution is presented in Appendix~\ref{sec:inf_calculation}.

\begin{table}
 	\centering
 	\caption { Table of results of the inference test}
 	\label{tab:table_inftest}
 	\begin{tabular}{lccr} 
 		\hline
 		models & Total & 68( \% )   & 95 (\% )\\
 		\hline
 		random models & 480    & 290 ($60.42 \pm 2.29$\%)    &439  ($94.67 \pm 1.25 $\%)\\
 		single model & 200    & 134 ($67.00 \pm 3.50$\%)     &190  ($95.00 \pm 1.50 $\%)\\

 		\hline
 	\end{tabular}
 \end{table}

In practice, we perform an inference test on a set of random models on the thermal grid to test the overall performance of our inference algorithm.
We pick 12 models and execute 40 realizations per model. 
The result shows that the true values are within the 1-$\sigma$ (68\%) interval for $60.42\pm2.29$~\% (290/480) of the time, 
and in the 2$\sigma$(95\%) interval for $94.67 \pm 1.25$~\% (439/480) of the time, 
while the upper and lower limits are given by the $\pm 1 \sigma_\text{bi}$ error for corresponding binomial distributions. 
In addition, we carry out a cross-validation test to ensure our emulators are not affected by over-fitting problem. 
Here we select a single model near the center of the parameter space ({$\log (T_0/\text{K})$} = 3.643, $\gamma=1.591$, and {$\log (\Gamma_{\mathHI{}} /\text{s}^{-1})$} = -13.458.), 
and exclude the model\footnote{
In practice we exclude all models with the same $T_0$ and $\gamma$
{$\log (T_0/\text{K})$} =3.643 and $\gamma=1.591$), 
since we mostly want to test the performance of the \bndist{} emulator on the $T_0$-$\gamma$ plane.} 
from the training dataset.
We train our emulators (both \bndist{} and \dndz) based on the new dataset,
and run 200 realizations of our inference method.
We observe that the true values are inside the
1$\sigma$ (68\%) interval for $67.00 \pm 3.50$\% (134/200) of the time,
and inside the 2$\sigma$ (95\%) interval for $95.0 \pm 1.50 $\% (190/200) of the time.
Results are presented in Table~\ref{tab:table_inftest}.
The overall performance indicates that our algorithm passes the inference\footnote{
Our inference method performs better when the model is close to the center of the grid. 
This might be because our emulators, both \ac{DELFI} and Gaussian process emulator, 
perform better at the center of the grid where the interpolation is more accurate.
Besides, our thermal grid has an irregular shape on the $T_0$-$\gamma$ plane, 
and might thus make the interpolation even harder
or distorted when there are no or only a few models around. 
Such a problem might be addressed by adding more simulation models, extending the thermal grid to make sure the region we are interested in always lies at the center of the grid.}.

In the end, to further demonstrate and elaborate on the effectiveness of our inference algorithm, 
we created a toy model, which involves entire inference pipeline, (in Appendix~\ref{sec:toymodel}) 
to test the whole inference algorithm under more controlled conditions,
where the toy \bndist{} is analytical, and the parameter dependence is known.
Here the toy \bndist~consists of a multivariate Gaussian distribution
parameterized by three mock parameters following the parameter dependence discussed in \S~\ref{sec:parameter}.
Moreover, these mock parameters also control the line density \dndz{} of the model based on the \dndz{} map generated by the Gaussian emulator from our Nyx simulation models (see Appendix~\ref{sec:toymodel} for more details). 
As a result of this toy model and also the inference test, we conclude that our inference algorithm is sound.

\section{Summary and Conclusions}
\label{3sec:discussion}

In this study, we have presented and evaluated our new method of measuring the thermal state $[T_0,\gamma]$ and the photoionization rate $\Gamma_{\mathHI{}}$ of the low redshift \ac{IGM} using its \bndist{} and absorber density \dndz{}.
We made use of a novel machine learning technique \ac{DELFI} to build a \bndist{} emulator and used a Gaussian process emulator to simulate the absorber density \dndz{}. 
We trained both emulators on a dataset generated 
from a set of Nyx simulations on a large parameter grid.
To test the performance of our inference algorithm under realistic conditions, 
we applied forward modeling techniques to model the noise and 
instrumental effects based on the HST \ac{COS} quasar spectra from \cite{Danforth2016}.
We showed using extensive tests that our inference method is proficient and reliable.  
Here we conclude by discussing the performance and summarizing the essential elements of our new algorithm. 

\begin{itemize}

\item We used mock datasets to simulate the measurement of the thermal state $[T_0,\gamma]$ of the low redshift \ac{IGM} from the full joint \bndist{}, for the first time taking the absorber density \dndz{} into account. The latter
enables us to constrain the photoionization rate $\Gamma_{\rm HI}$,
since only the shape of the \bndist{} is insensitive to this parameter (see Fig.\ref{fig:shiftpdf}). 
We also confirm that the \dndz{} term we introduced is consistent with our inference based on the \bndist{} alone, and improves the performance of our inference method (see Appendix \ref{sec:2dresult}). 

\item Our new inference method successfully recovers thermal parameters of models 
from the Nyx simulation with small uncertainties (in our example, $\sigma_{\log T_0} \sim 0.08$~dex, $\sigma_\gamma \sim 0.06$, and $\sigma_{\log \Gamma_{\mathHI{}}} \sim 0.07$~dex),
using a relatively small dataset with $\Delta z =2.316$. 
Furthermore, these results are obtained under realistic conditions as we forward-model the observational effects and noise from the \cite{Danforth2016} low-$z$ \ac{COS} quasar spectra while setting the size of our mock datasets to be the same as the observational dataset (i.e. having the same total pathlength $\Delta z_{\text{ob}}$). 
Considering all these factors, the accuracy and sensitivity we attained in this study should be achievable when our inference method is applied to real observational data, 
making it a powerful tool for studying the \lyaf{}. 

\item Our algorithm passes the inference test (see \S\ref{sec:Inference_test}),
indicating that our approximation and emulation/interpolation are reliable.
We also demonstrate the robustness of our inference method by testing the entire inference pipeline,
including emulation and interpolation procedures on a toy model under better-controlled conditions(see Appendix \ref{sec:toymodel}).

\item The \bndist{} (\ac{DELFI}) emulator successfully emulates both the 2D \bndist s and 1D marginalized distributions of $b$ and $N_{\mathHI{}}$. 
We find that the 2D \bndist{} shifts upward (towards higher $b$ values) 
with increasing $T_0$ and $\gamma$, 
while larger $\gamma$ also tilts up the low-$b$ cut off.
We explain these effects qualitatively in section \S~\ref{sec:parameter} 
and show that they are consistent with previous work.

\end{itemize}

Moreover, previous work \citep[][]{Viel2017,Gaikwad2017,Nasir2017} reported a discrepancy in the
1D marginalized $b$ distribution for low redshift \ac{IGM} between the observation and current simulations,
implying the existence of additional heating
or turbulence that is stronger than expected \citep[][]{Bolton2021}. 
{While these works mainly focus on 1D marginalized distributions of $b$ and \ac{CDDF}, 
our new inference algorithm, which successfully emulate both 1D marginalized and 2D joint \bndist{}, 
would allow us to investigate such problem using the joint distribution together with \dndz{} statistics.}
We aim to investigate this problem by applying our inference method to observational data in future works, 
which we expect would provide an accurate measurement of the thermal state of the low-$z$ \ac{IGM} and possibly solve this discrepancy.
In addition, 
{we also look forward to applying our method to 
other recent cosmological galaxy formation simulations 
like Illustris (TNG) \citep{Genel2014,Weinberger2017}, 
to study the effect of feedback on the \lyaf{}~which is
not yet completely understood~\citep[see for e.g,][]{Gurvich2017, Christiansen2020, Blakesley22}.} 

\section*{Acknowledgements}

We thank the members of the ENIGMA\footnote{http://enigma.physics.ucsb.edu/}, 
Siang Peng Oh, Timothy Brandt, and K.G. Lee for helpful discussions and 
suggestions. Thanks also to Ilya Khrykin for useful feedback as well as contributions to the 
inference code.

Calculations presented in this paper used the hydra and draco clusters
of the Max Planck Computing and Data Facility (MPCDF, formerly known
as RZG). MPCDF is a competence center of the Max Planck Society
located in Garching (Germany).
This research also used resources of the National Energy Research Scientific Computing Center (NERSC),
a U.S. Department of Energy Office of Science User Facility located at Lawrence Berkeley National Laboratory, operated under Contract No. DE-AC02-05CH11231. 
In addition, we acknowledge Partnership for Advanced Computing in Europe (PRACE) for awarding us access to JUWELS hosted by JSC, Germany.

Justin Alsing was supported by research project grant Fundamental Physics from Cosmological Surveys funded by the Swedish Research Council (VR) under Dnr 2017-04212.

\section*{Data Availability}

The simulation data and analysis code underlying this article will be shared on reasonable request to the corresponding author.

%%%%%%%%%%%%%%%%%%%%%%%%%%%%%%%%%%%%%%%%%%%%%%%%%%



\bibliographystyle{mnras}


\bibliography{references.bib}

%%%%%%%%%%%%%%%%%%%%%%%%%%%%%%%%%%%%%%%%%%%%%%%%%%

\appendix

\section{Inference without absorber density}
\label{sec:2dresult}

  \begin{figure*}
 \centering
    \includegraphics[width=0.8\linewidth]{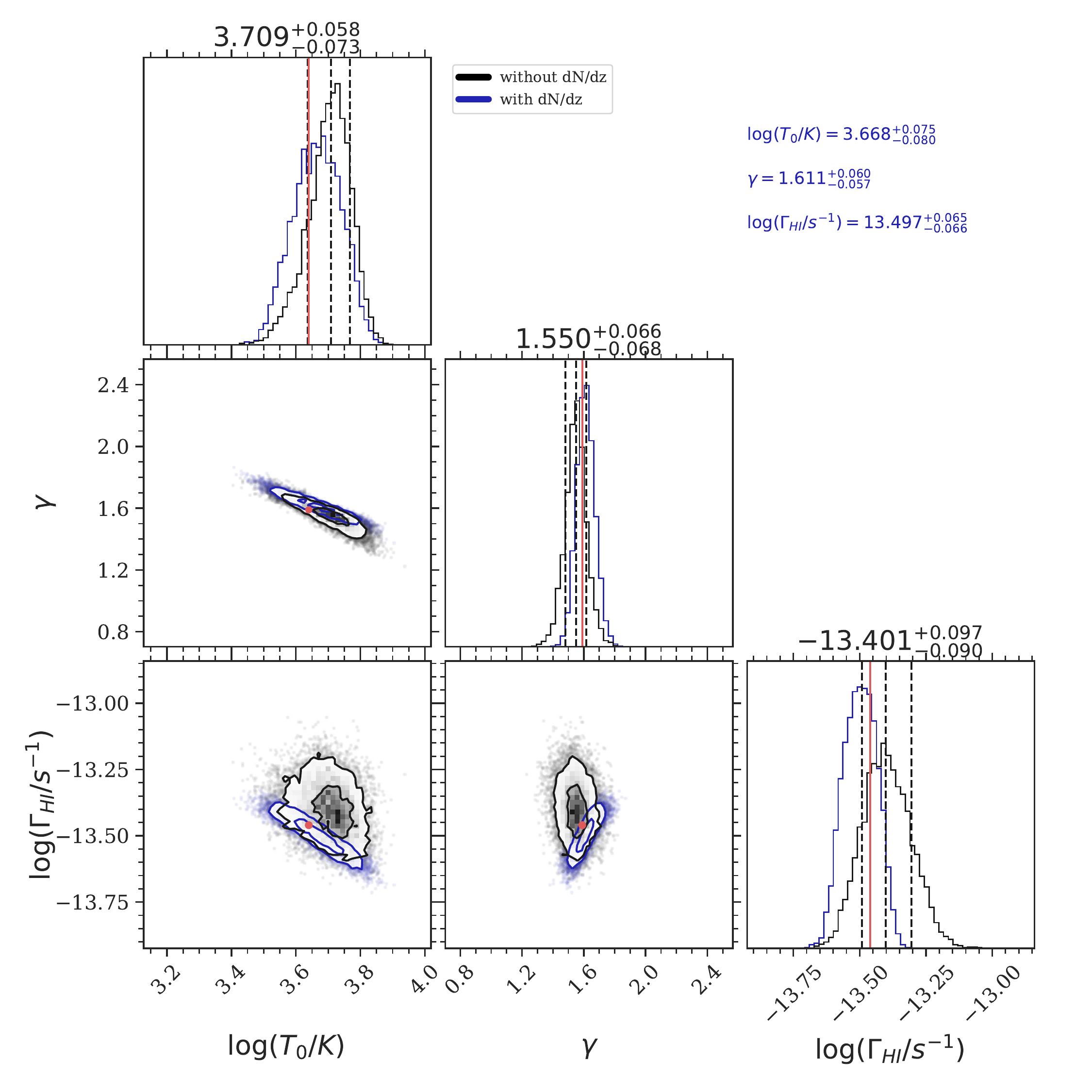}
  \caption{MCMC posterior (black) for the Nyx model discussed in \S\ref{sec:3dresult} 
  based on the likelihood function without the absorber density Eq.(~\ref{eq:eq_nodndz}). Projections of the true model is shown as red dot. Inner(outer) contours represents the projected 2D 1(2)-sigma interval. The parameters of true model are indicated by red lines in the marginal distributions, while the dashed black lines indicates the 16, 50, and 84 percentile values of the posterior. The true parameters are: {$\log (T_0/\text{K})$} = 3.643, $\gamma=1.591$ and {$\log (\Gamma_{\mathHI{}} /\text{s}^{-1})$} = -13.458. In comparison, the posterior obtained using Eq.(\ref{eq:log-likelihood}), which takes into account the \dndz{}, is shown in blue,
  while the medians of the posterior are shown in blue on the top right.  }
   \label{fig:corner_Nyx_NOdndz}
\end{figure*}  

In this section we provide more details about the inference without using the absorber density.
In such  a case, the likelihood function would simply be the first term of Eq.(\ref{eq:likelihood2}),
i.e.
\begin{equation}
\ln \mathcal{L}= \sum_{i=1}^{n} \ln P(b_i, N_{\mathHI{},i}).
\label{eq:eq_nodndz}
\end{equation}
Such likelihood function is evaluated based on our \bndist{} emulator solely. To make better comparison, we use the same mock dataset and training dataset as used in \S\ref{sec:3dresult}. The MCMC posterior is given in Fig.\ref{fig:corner_Nyx_NOdndz}, where we obtain $\log (T_0\text{K}) = 3.709^{+0.058}_{-0.073}$, $\gamma=1.550^{+0.066}_{-0.068}$ and  $\log (\Gamma_{\mathHI{}} /\text{s}^{-1})=13.401^{+0.097}_{-0.090}$ from the marginalized distributions, whereas the true parameters are: $\log (T_0\text{K}) = 3.643$, $\gamma=1.591$ and $\log (\Gamma_{\mathHI{}} /\text{s}^{-1}) = 13.458$. 
In comparison, the posterior obtained using Eq.(\ref{eq:log-likelihood}),
which takes into account the \dndz{}, is shown in blue in Fig.\ref{fig:corner_Nyx_NOdndz}.
As we show here, the two inference results are coherent,
but our modified inference algorithm (green posteriors) perform better.
By implementing the \dndz{} feature, 
our modified inference algorithm provides more accurate results in both  $T_0$ and $\Gamma_{\mathHI{}}$, and reduce the uncertain in $\Gamma_{\mathHI{}}$ significantly.
Furthermore, the inference without absorber density dose not pass the inference where the true model falls in the 1-$\sigma$ (68\%) interval for about 50\% of the time. 

In short, by employing the absorber density we not only evidently reduce the uncertainty in $\Gamma_{\mathHI{}}$ but also increase the accuracy in other parameters since the modification adds more information to the Bayesian analysis by matching the absorber density.

\section{Inference test likelihood calculation}
\label{sec:inf_calculation}

  \begin{figure*}
 \centering
    \includegraphics[width=0.9\linewidth]{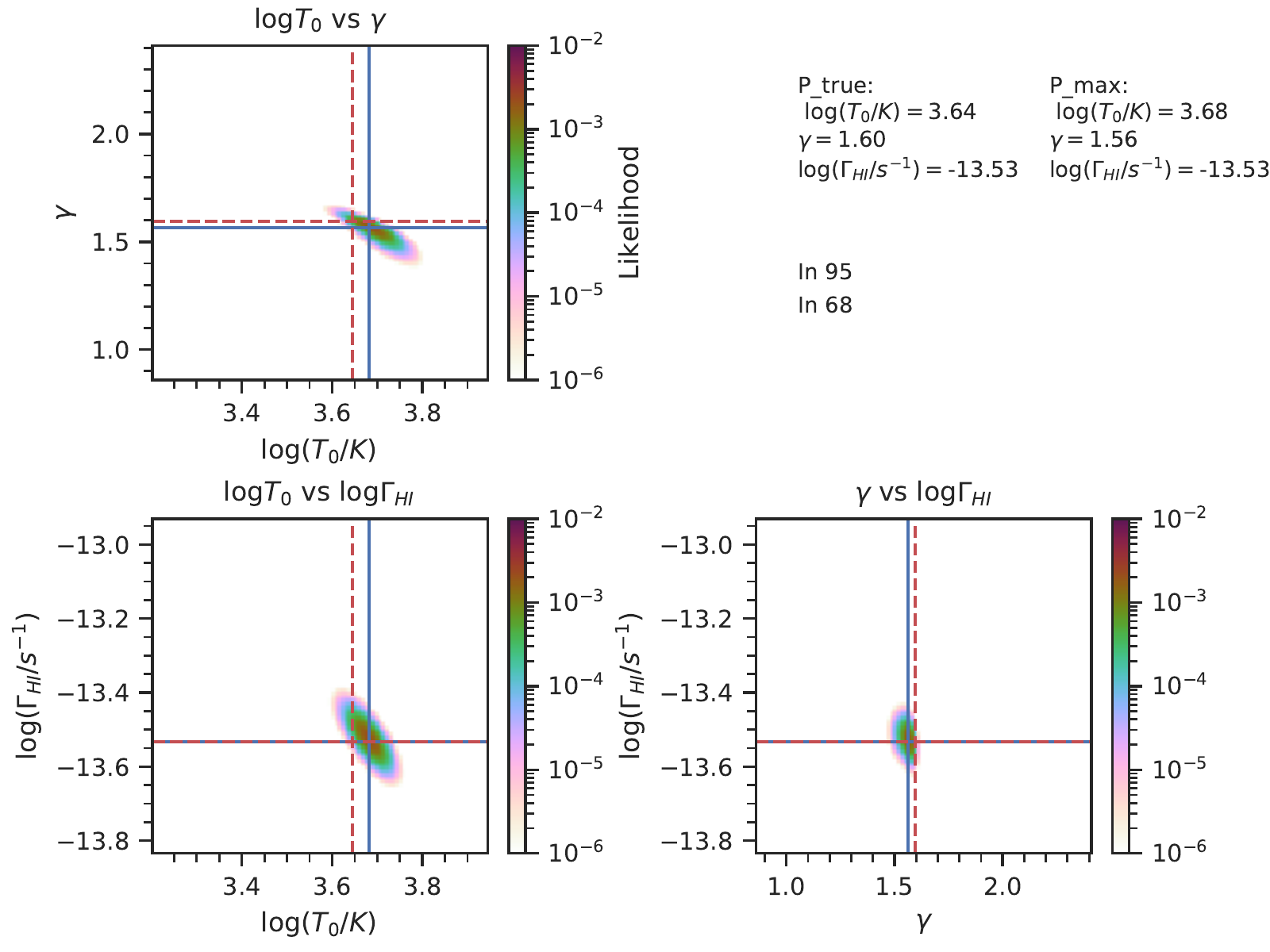}
  \caption{ Example of the distribution of the likelihood function sliced at the location of the true parameters ($P_{\text{true}}$, 
  indicated by red dashed lines). The parameters corresponding to the maximum likelihood model $P_{\text{max}}$  
  are indicated by blue solid lines.
  Values of both $P_{\text{true}}$ and $P_{\text{max}}$ are given in the up right.
  Calculation implies that the true parameters are in both the effective 1$\sigma$ 68\%) and 2$\sigma$ (95\%) intervals.}
  \label{fig:inftest_1}
\end{figure*} 

 To calculate contours of cumulative probability distribution with high dimensionality is challenging in computation power. 
In our case, the parameter grid size is $100^3$ and we have to compute the probability density function \bnpdfG{} many hundreds times 
(i.e. the number of lines in the data set) to evaluate the likelihood function on a single point on the parameter grid (see Eq.~\ref{eq:log-likelihood}).
However, due to the structure of the $b$-$N_{\rm HI}$ PDF calculated by our DELFI emulator, 
we are able to save time by computing the likelihood function on the whole grid simultaneously, 
with help of vector operations implemented in \texttt{python}, 
though such treatment requires reconstruction of the likelihood function and needs extra amounts of memory. 
In comparison, our code is much faster than the \ac{MCMC} prescription which would require a very long chain to interpolate the likelihood function on the whole grid to achieve the same precision. 
An example of the distribution of the likelihood function is shown in Fig.\ref{fig:inftest_1}. 

\section{Toy model}
\label{sec:toymodel}

 \begin{figure}
 \centering
    \includegraphics[width=0.8\columnwidth]{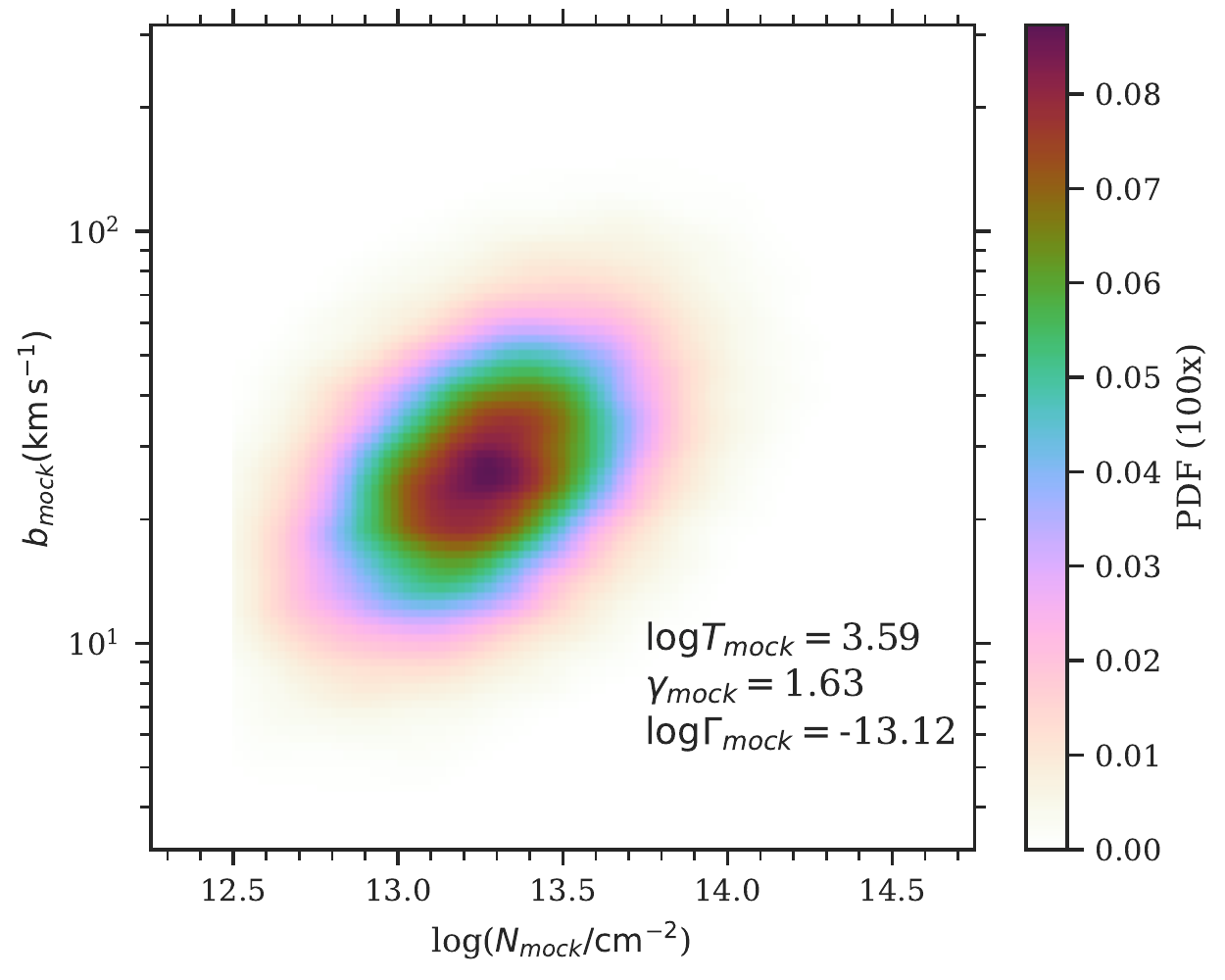}
  \caption{ The KDE based PDF of \bndist~of one of the toy models  which is a 2D Gaussian distribution parameterized by $T_{\text{mock}}$, $\gamma_{\text{mock}}$ and $\Gamma_{\text{mock}}$ in analogy with thermal parameters $T_0$, $\gamma$ , $\Gamma_{\mathHI{}}$. The parameters of the toy model is shown in the right bottom corner of the plot. For illustration purposes, values of pdf are multiplied by 100 in the color bar.}
  \label{fig:toymodel}
\end{figure} 

 \begin{figure*}
 \centering
    \includegraphics[width=0.85\linewidth]{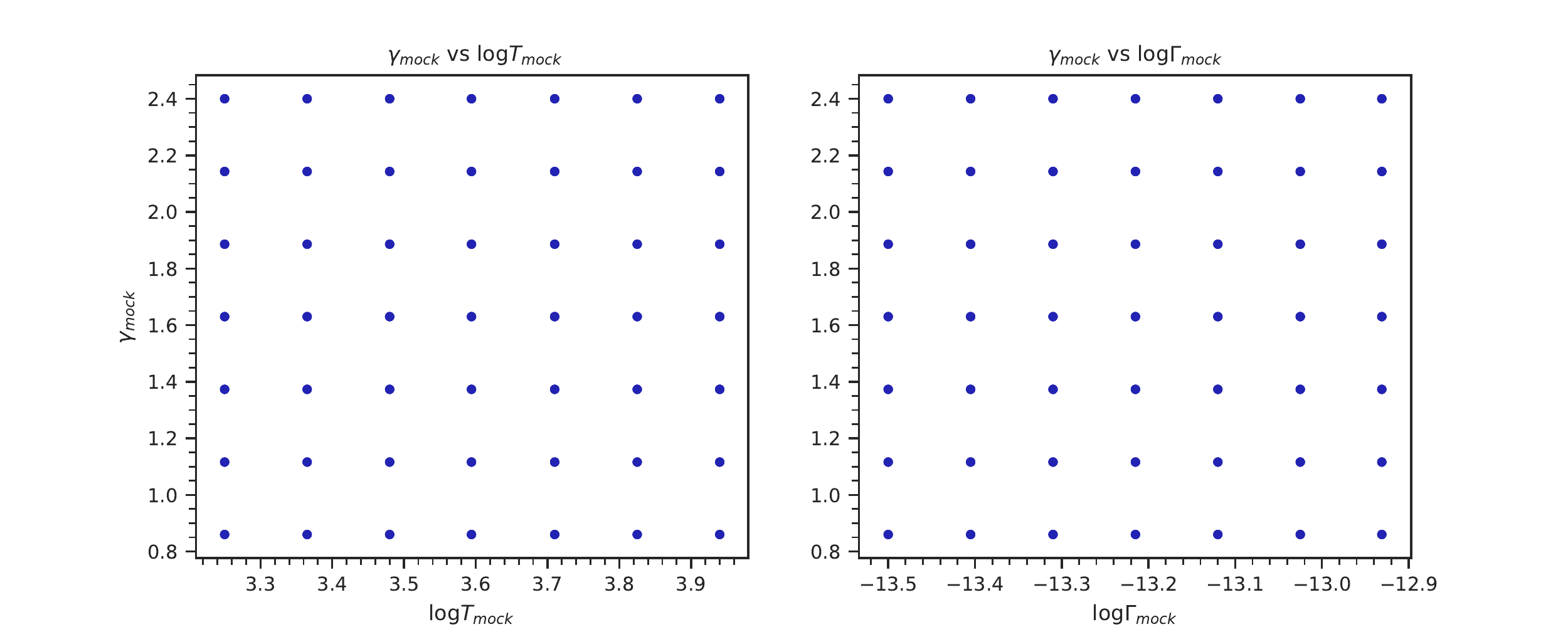}
  \caption{The thermal grid used in our toy model. The left-hand panel is the $\gamma$ - $T_0$ grid and the right-hand panel is $\gamma$ - $\Gamma_{\mathHI{}}$ slice showing the 7 $\Gamma_{\mathHI{}}$ values we have for each point on the 2D $\gamma$ - $T_0$ grid.}
  \label{fig:toy_grid}
\end{figure*}

 \begin{figure}
 \centering
    \includegraphics[width=\columnwidth]{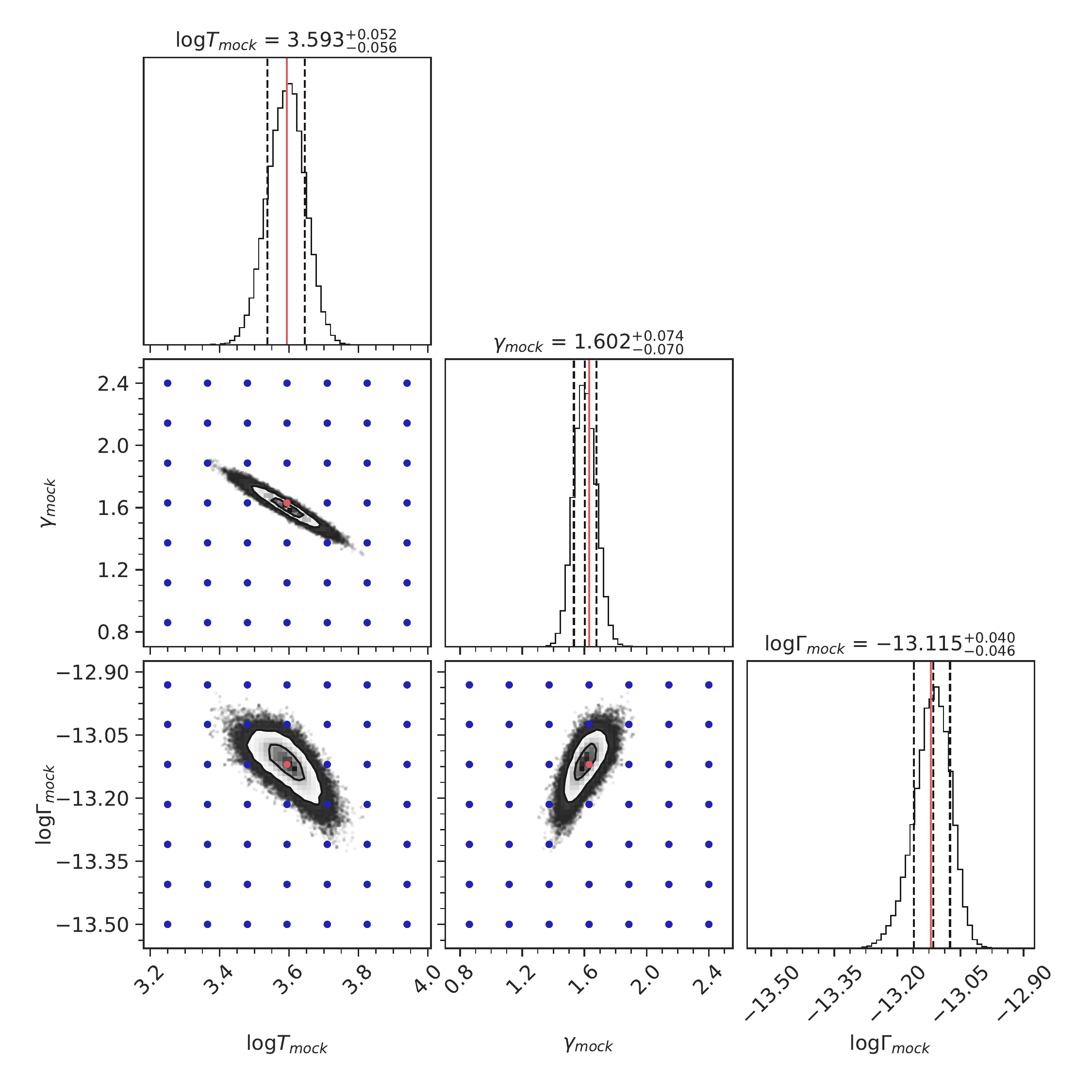}
  \caption{\ac{MCMC} posterior for the fit of the \bndist{} from one of the toy models (absorbers shown as black points in Fig.\ref{fig:fit_toy}) using the likelihood function (Eq.~\ref{eq:likelihood}) from \ac{DELFI} and our Gaussian emulator (see \S~\ref{3sec:log-likelihood}). Projections of the thermal grid used for generating models are shown as blue circles.  Inner(outer) black contour represents the projected 2D 1(2)-sigma interval. The parameters of true model are indicated by red lines in the corner plot, while the dashed black lines indicates the 16, 50, and 84 percentile value of the posterior. The true parameters are: {$\log T_{\text{mock}} = 3.59$, $\gamma_{\text{mock}}=1.63$ and $\log \Gamma_{\text{mock}} = -13.12$.  }}
  \label{fig:corner_toy}
\end{figure}

 \begin{figure}
 \centering
    \includegraphics[width=\columnwidth]{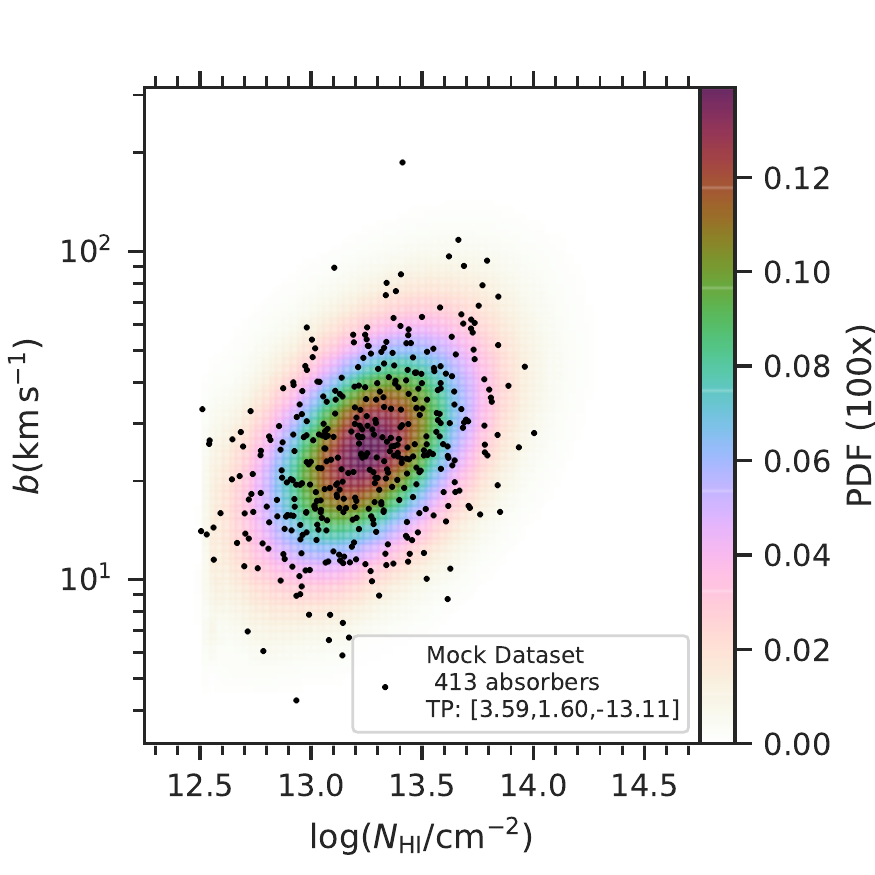}
  \caption{The `best fit' model \bndist~for the Gaussian toy model emulated by DELFI. It is emulated based on the best-fit parameters (median values of the marginalized MCMC posterior), which is shown in the right bottom corner of the plot. The true parameters are: {$\log (T_{\text{mock}}$} = 3.59,{ $\gamma_{\text{mock}}=1.63$} and {$\log (\Gamma_{\text{mock}}$} = -13.12. For illustration purposes, values of pdf are multiplied by 100 in the color bar.}
  \label{fig:fit_toy}
\end{figure} 

To verify the performance of our emulators in a clean environment,
we build a toy model with a mock data set which roughly simulates the behavior of our real model. 
Here the toy \bndist{}s consist of 2D Gaussian distributions parameterized by $T_{\text{mock}}$ and $\gamma_{\text{mock}}$, $\Gamma_{\text{mock}}$ in analogy with thermal parameters $T_0$, $\gamma$ and $\Gamma_{\mathHI{}}$. 
Here we follow the parameter dependence discussed in \S\ref{sec:parameter}, 
i.e. both $T_{\text{mock}}$ and {$\gamma_{\text{mock}}$} sets the $y$-axis location of the center of the Gaussian,
while $\gamma_{\text{mock}}$ also sets the tilted angle of the Gaussian,
and the $\Gamma_{\text{mock}}$ controls the density of data points for each model, 
in analogy with the $\Gamma_{\mathHI{}}$ which determines the absorber density \dndz{}. 
{For convenience, we set these mock parameters to be dimensionless.} 
We tune these parameters in a way that the `\bndist{}' of our toy model falls roughly in the same range  as the Nyx simulation, 
and we adopt absorber density emulated by our \dndz{} emulator based on our Nyx simulations, 
so that the mock \dndz{} follows the relationship between thermal parameters and absorber density in our Nyx simulation. 
We in total generate 7x7x7 = 343 (see Fig.\ref{fig:toy_grid}) models spanning the thermal grid. 
An example of the \bndist~of a toy model is shown in Fig.\ref{fig:toymodel}, which is generated based on the Kernel Density Estimation (KDE) of the mock dataset using a smoothing bandwidth $(\sigma_{\log N_\text{mock}}, \sigma_{\log b_\text{mock}})= (0.08,0.32)$. Such choice of bandwidth is taken from \citet{Hiss2019}.

For each toy model with different mock thermal parameters, 
we first generate a set of 2000 `imaginary' pathlength $\Delta z_{i}$, 
each of which equals to a randomly chosen observation spectra in \citet{Danforth2016} low-z \lya{} dataset (i.e. for each model we generate a set of 2000 $\Delta z_{i}$ but without actual spectra). 
For each `imaginary' pathlength $\Delta z_{i}$
we generate a set of mock `$b$-$N$' pairs (lines), sampling from the \bndist{}, 
while the number of lines $N_i$ follows a Poisson distribution $\text{Pois}(\lambda_i)$ with Poisson rate $\lambda_i =\Delta z_i \times\left({\text{d} N}/{\text{d} z}\right)_{\rm model}$, 
where the $\left({\text{d} N}/{\text{d} z}\right)_{\rm model}$ is the absorber density of that model.
The total number of lines for the model is thus 
$N_\text{tot} = \sum_i^{2000} N_{i}$.
At this point we obtain a training dataset with the same structure as the one described in \S\ref{sec:emu}, which consists of `$b$-$N$' pairs labeled by thermal parameters.
We then train the DELFI (\bndist) and Gaussian (\dndz{}) emulators based on the above dataset, 
and test our whole inference algorithm on the toy model following the prescription given in \S\ref{sec:3dresult}. An example of the inference result is shown below, including the  MCMC posteriors (Fig.\ref{fig:corner_toy}) and the 'best fit' \bndist{} recovered from mock dataset (Fig.\ref{fig:fit_toy}). As a comparison, the KDE based PDF of the \bndist{} of the model is shown in Fig.\ref{fig:toymodel}.

In the end, we perform inference test on our toy model for both 3D and 2D (without $\Gamma_{\mathHI{}}$) models to test the robustness of our whole inference pipeline following the method discussed in \S\ref{sec:Inference_test},
and the results are given in table \ref{tab:toy_inftest}, 
showing that our inference algorithm passes the inference test perfectly for an idealized model. 
Moreover, the inference on toy model of \bndist{} performs 
slightly better than on Nyx simulation (see Appendix \ref{sec:toymodel}). 
The reason could be that the toy model \bndist s are 2D Gaussian distributions that solely depends on the thermal parameters $T_{\text{mock}}$ and $\gamma_{\text{mock}}$, which is equivalent to say that the \bndist{} fully preserved the thermal information of the IGM, however, in the Nyx simulation the \bndist s are affected by the complex astrophysical processes in the diffuse IGM, resulting in the loss of the thermal information.

 \begin{table}
 	\centering
 	\caption { Table of results of the inference test for the toy model}
 	\label{tab:toy_inftest}
 	\begin{tabular}{lccr} 
 		\hline
 		models & Total & 68( \% )   & 95 (\% )\\
 		\hline
 		3D toy model & 240    & 165 ($68.75 \pm  2.92$\%)    &225  ($93.75 \pm 1.67$\%)\\
 		2D toy model & 300   & 199 ($66.33 \pm   2.67$\%)    &284  ($94.67 \pm 1.33 $\%)\\

 		\hline
 	\end{tabular}
 \end{table}

Combining all results shown above, we conclude that our inference algorithm is able to recover the mock parameters with extraordinary accuracy under idealized condition, and our entire pipeline including \bndist{} emulation, \dndz{} emulation, likelihood function and inference pipeline is robust. 

\section{Convergence test}
\label{sec:convergence}

 \begin{figure*}
 \centering
    \includegraphics[width=0.85\linewidth]{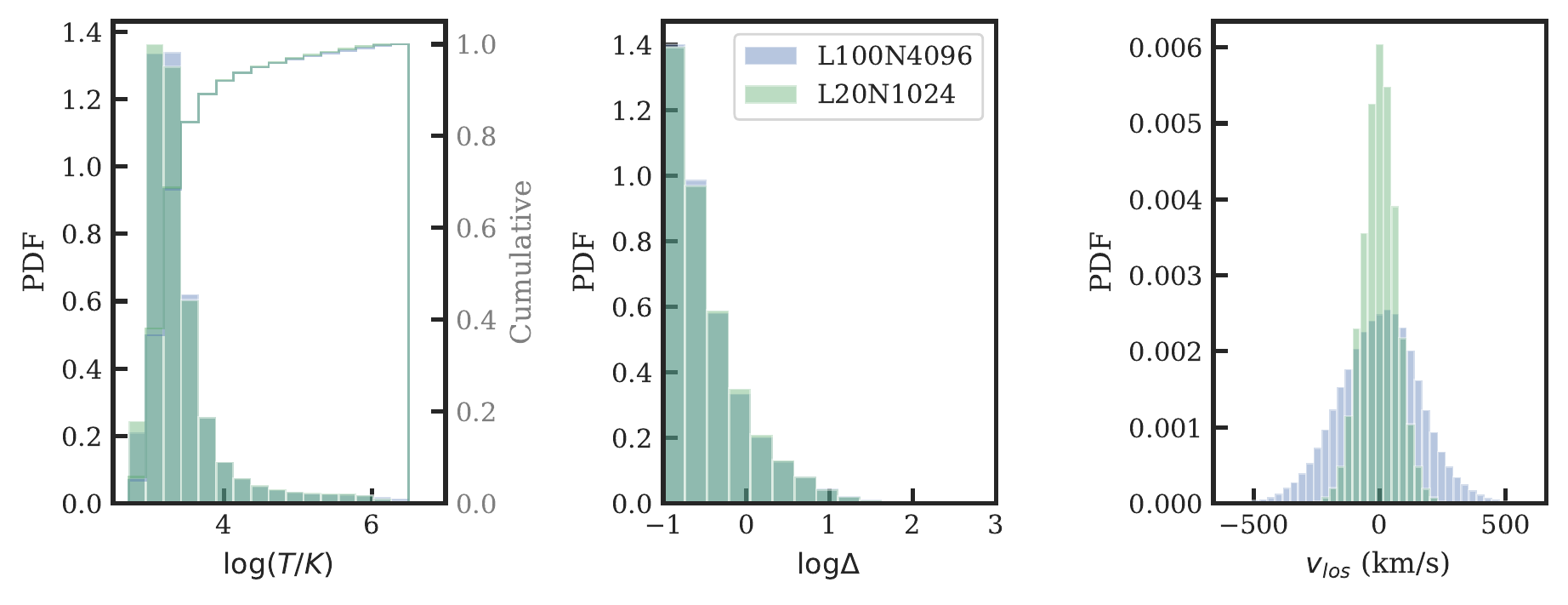}
  \caption{ From left to right, the 1D marginalized distribution of the temperature $T$, overdensity $\Delta$, and velocity along line-of-sight $v_\text{los}$. The unfilled histogram in the left most panel shows the CDF of the temperature distribution. The \emph{L100N4096} box are shown in blue, while the \emph{L20N1024} box are shown in green.}
  \label{fig:rho_T_converge}
\end{figure*}

 \begin{figure*}
 \centering
    \includegraphics[width=0.85\linewidth]{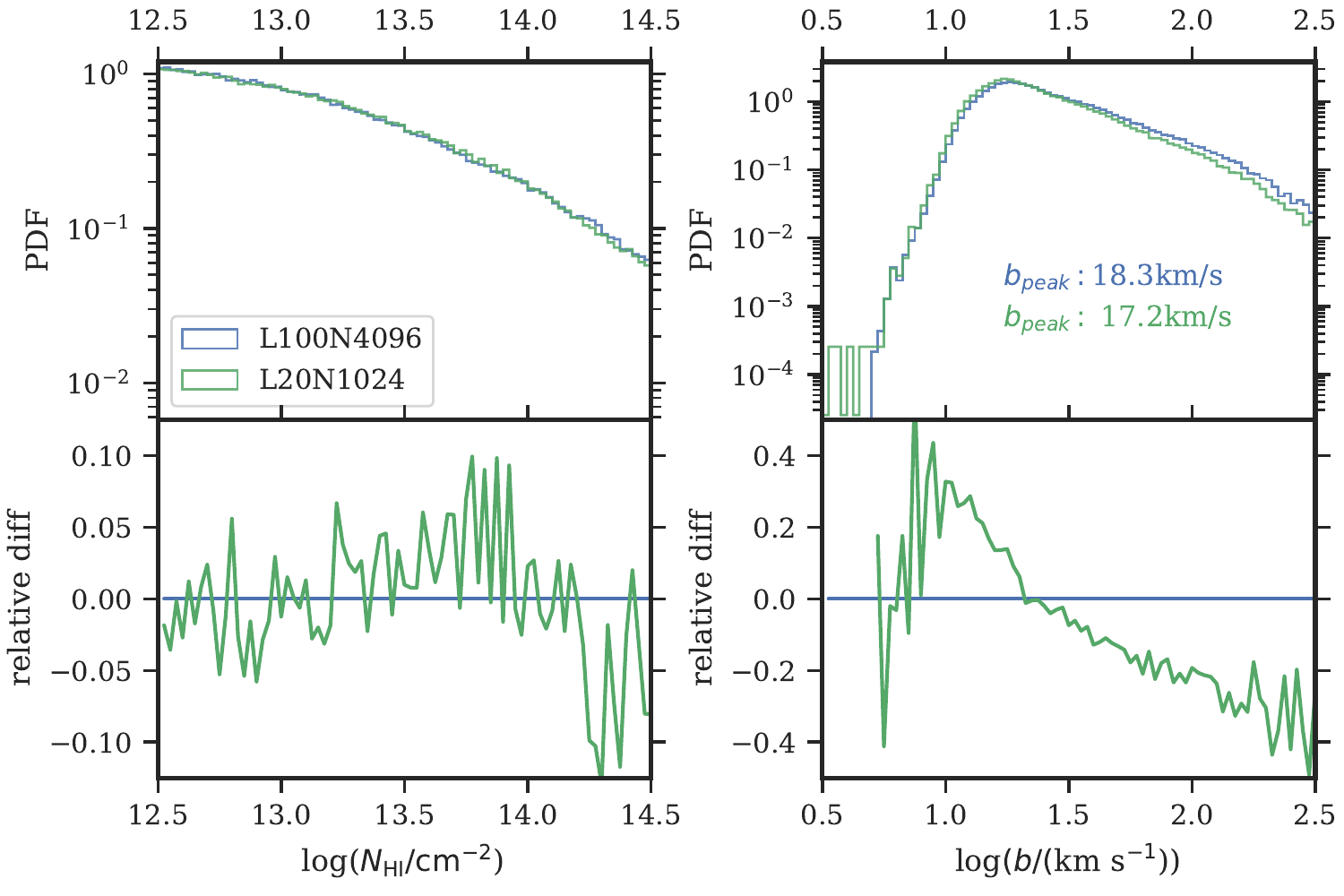}
  \caption{ The 1D marginalized $N_{\rm HI}$(left) and $b$(right) parameters of the two simulations. The relative differences are shown in the bottom panels. The \emph{L100N4096} box is shown in blue, while the \emph{L20N1024} box is shown in green. The peak of $b$ parameters are given in the text.}
  \label{fig:1d_bN_converge}
\end{figure*} 

 \begin{figure*}
 \centering
    \includegraphics[width=1\linewidth, keepaspectratio]{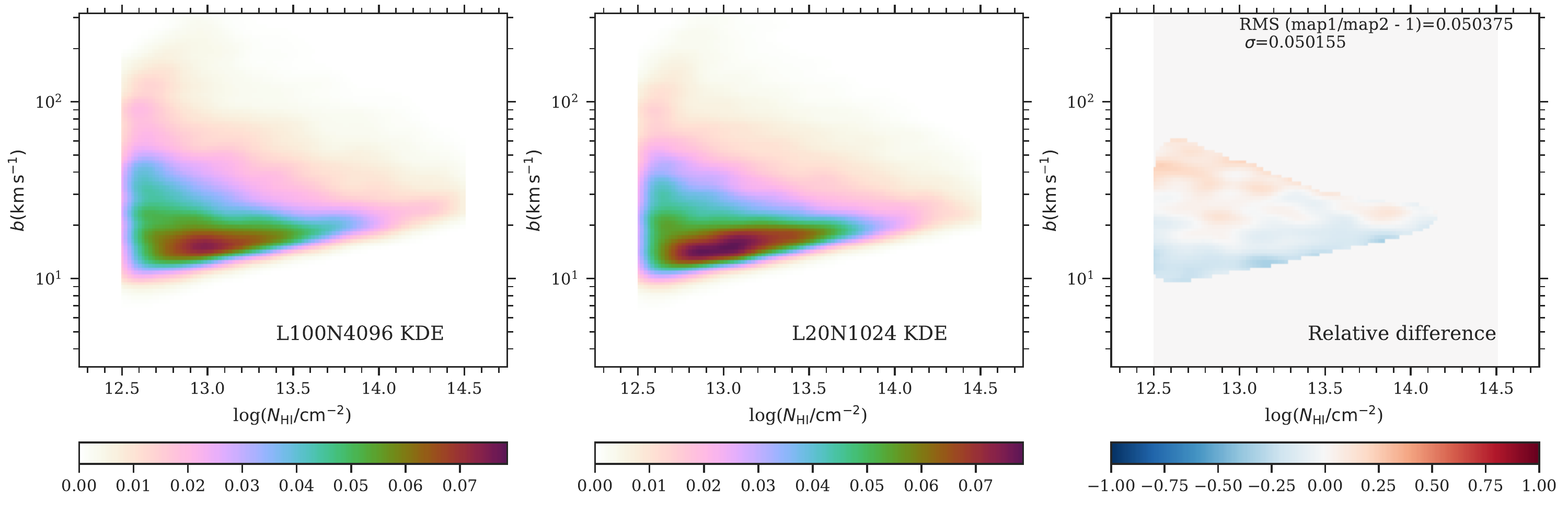}
  \caption{ The 2D KDE maps of \bndist{}s for simulations
  \emph{L100N4096} (left-hand panel) and \emph{L20N1024}  (middle panel). 
  The right-hand panel shows the relative difference $\Delta_{P}=(P_{L100}/P_{L20} -1)$.  
  To avoid division by zero, we apply a threshold
  and only include regions integrating up to 75\% for $P_{L20}$.
  The KDE maps are made from 20000 data points for each simulation,
  and the RMS and
  standard deviation of 
  of the relative difference map are given in the right-hand panel. 
  Details of the calculations are given in the text.}
  \label{fig:bN_KDE_converge}
\end{figure*} 

{\cite{Lukic2015} demonstrated that 
the $b$ parameter of \lyaf{} is sensitive to the simulation resolution, and its distribution converges for simulation finer than \emph{L10N512} 
simulation 
(i.e.,
box size $L =10 h^{-1}$ Mpc and 
$N=512^3$ dark matter particles and baryon grids which gives the 
resolution of 20 h$^{-1}$ kpc)
while the box size
itself does not 
affect line parameters of the \lyaf{}. 
Whereas above mentioned tests are done at redshift $\sim 3$,
it is worthy to further investigate impact of the boxsize 
and resolution 
of the simulation on the \lyaf{} at lower redshifts, 
since the nonlinear evolution 
at low redshift 
can affect the \lyaf.} 

{Here, we perform a convergence test at redshift  $z = 0.5$ to 
to check if 
our results are independent of the simulation box-size
at low redshift.
To test the convergence 
we use two Nyx boxes; 
\emph{L20N1024} (box-size = 20 $h^{-1}$ Mpc, 
$N=1024^3$ dark matter particles and baryon grids i.e
resolution of 20 $h^{-1}$ kpc), and \emph{L100N4096}
(box-size  100 $h^{1-}$ Mpc and $N=4096^3$ dark matter particles and 
baryon grids, resolution of 24 $h^{-1}$ kpc).
These two simulation boxes are ran following the same procedures given in section~\S\ref{3sec:simulations}. 
In Fig.~\ref{fig:rho_T_converge}, we plot the temperature $T$, overdensity $\Delta$, and velocity along line-of-sight $v_{los}$ of 
these two simulations. We can see the distributions of $T$ and $\Delta$ are alike 
for both while the small box \emph{L20N1024} simulation has much smaller line-of-sight velocity. This is expected 
since line-of-sight velocities are dominated by the large scale modes that exist only in the large box simulations. However, these large 
velocities are because of bulk motion and therefore
do not affect the parameters of the \lyaf{} lines.}

{For both simulations, 
we follow the forward modeling and line fitting procedures discussed in Section~\S\ref{3sec:simulations},
except that here we use a Gaussian LSF with 
fixed resolution R=3.5 km/s and assume a SNR=100.
Such choices of resolution and SNR assure that the \lyaf{} are fully resolved and the box-size effect are independent of resolution and instrument.
For both simulations, we use the photoionization rate $\log \Gamma_{\mathHI{}} (\text{s}^{-1}) = -13.308$. 
The 1D marginalized distributions of 
Doppler parameter $b$ and column density $N_{\mathHI{}}$ 
of both simulations are presented in Fig.~\ref{fig:1d_bN_converge}. 
The $N_{\mathHI{}}$ distribution of the two simulations are in excellent agreement with each other, with the relative difference  $\Delta_{P(N)} < 10\%$.
The $b$ parameter have very similar distributions for both simulations, 
where the two distributions agree with each other near the peak, with relative difference $\Delta_{P(b)} < 25\%$, and the difference increases as $\log b$ becomes smaller than $1.0$ or larger than $2.0$, which however have very small contribution in the total cumulative distribution.
The peak values of the $b$ parameter for both simulations are given in Fig.~\ref{fig:1d_bN_converge}, where the $b$ distributions give $b_\text{peak}$ = 18.3 km/s and 17.2km/s for \emph{L100N4096} and \emph{L20N1024} simulations respectively.
We count the \dndz{} for both simulations,
\emph{L20N1024} gives \dndz = 750,  and 
\emph{L100N4096} gives \dndz = 700. The difference in \dndz{} is about 7\%.
Furthermore, we plot the 2D \bndist{} 
in Fig.~\ref{fig:bN_KDE_converge} for both simulations. 
These are 2D KDE maps each generated by 20000 data points
collected from the \bn{} dataset following the procedures described in
Section~\S\ref{3sec:simulations}. In the right most panel of 
Fig.~\ref{fig:bN_KDE_converge}, we plot the relative difference of the 
KDE map, given by $\Delta_{P}=(P_{L100}/P_{L20} -1)$, where the $P_{L100}$ and $P_{L20}$ stand for the KDE for
\emph{L100N4096} and \emph{L20N1024} simulations respectively. 
To avoid division by zero, we apply a small threshold and only include regions 
with $P_{L20} > P_\text{TH}$, 
where $\int_{P_\text{TH}}^\infty P \text{d}P =75\%$.
We quantify the overall relative difference 
by calculating the root mean square and standard deviation of the $\Delta_{P}$. 
As shown in Fig.~\ref{fig:bN_KDE_converge}, the 
relative differences in the 2D \bndist{} are small and only about 5\%. 
Therefore we conclude, even at $z\sim 0.5$ box-sizes do not affect the
parameters of \lyaf~significantly.}

%%%%%%%%%%%%% SECTION %%%%%%%%%%%%%
%\section{appendix}
%\label{app:}

\begin{acronym}
	\acro{AGN}{active galactic nuclei}
	\acro{CDDF}{column density distribution function}
	\acro{CMB}{Cosmic Microwave Background}
	\acro{COS}{\emph{Cosmic Origins Spectrograph}}
	\acro{DELFI}{density-estimation likelihood-free inference}
	\acro{DM}{dark matter}
	\acro{DLA}{damped Ly$\alpha$}
	\acro{GP}{Gaussian process}
	\acro{HIRES}{High Resolution Echelle Spectrometer}
	\acro{HST}{\emph{Hubble Space Telescope}}
	\acro{IGM}{intergalactic medium}
	\acro{KDE}{Kernel Density Estimation}
	\acro{KODIAQ}{Keck Observatory Database of Ionized Absorbers toward QSOs}
	\acro{LD}{least absolute deviation}
	\acro{LLS}{Lyman limit systems}
	\acro{LS}{least squares}
	\acro{LSF}{line spread function}
	\acro{MCMC}{Markov chain Monte Carlo}
	\acro{MW}{Milky Way}
	\acro{NDE}{neural density estimators}
	\acro{PCA}{principal component analysis}
	\acro{PDF}{probability density function}
	\acro{PKP}{\ac{PCA} decomposition of \ac{KDE} estimates of a \ac{PDF}}
	\acro{QSO}{quasi-stellar objects}
	\acro{SNR}{signal-to-noise ratio}
	\acro{STIS}{\emph{Space Telescope Imaging Spectrograph}}
	\acro{TDR}{temperature-density relation}
	\acro{THERMAL}{Thermal History and Evolution in Reionization Models of Absorption Lines}
	\acro{UV}{ultraviolet}
	\acro{UVB}{ultraviolet background}
	\acro{UVES}{Ultraviolet and Visual Echelle Spectrograph}
	\acro{WHIM}{Warm Hot Intergalactic Medium}
\end{acronym}

%%%%%%%%%%%%%%%%%%%%%%%%%%%%%%%%%%%%%%%%%%%%%%%%%%

% Don't change these lines
\bsp	% typesetting comment
\label{lastpage}
\end{document}